\documentclass[12pt]{iopart}

\usepackage[utf8]{inputenc}
\usepackage{amsfonts}
\usepackage{graphicx}
\usepackage{hyperref}   
\usepackage{color}
\usepackage{hyperref}
\hypersetup{
 	colorlinks=true,
	linkcolor=blue,
	citecolor=blue
}

\begin{document}

\title{Group theoretical and topological analysis of the quantum spin Hall effect in silicene}
\author {F. Geissler \dag, J. C. Budich \ddag, B. Trauzettel \dag}
 \address{\dag\ Institute for Theoretical Physics and Astrophysics,
 University of W\"urzburg, 97074 W\"urzburg, Germany}
 \address{\ddag\ Department of Physics, Stockholm University, Se-106 91 Stockholm, Sweden}

\date{\today}

\begin{abstract}
Silicene consists of a monolayer of silicon atoms in a buckled honeycomb structure. It was recently discovered that the symmetry of such a system allows for interesting Rashba spin-orbit effects. A perpendicular electric field is able to couple to the sublattice pseudospin, making it possible to electrically tune and close the band gap. Therefore, external electric fields may generate a topological phase transition from a topological insulator to a normal insulator (or semimetal) and vice versa. The contribution of the present article to the study of silicene is twofold: First, we perform a group theoretical analysis to
systematically construct the Hamiltonian in the vicinity of the $K$ points of the Brillouin zone and find an additional, electric field induced spin-orbit term, that is allowed by symmetry. Subsequently, we identify a tight binding model that corresponds to the group theoretically derived Hamiltonian near the $K$ points.
Second, we start from this tight binding model to analyze the topological phase diagram of silicene by an explicit calculation of
the $\mathbb Z_2$ topological invariant of the band structure. To this end, we calculate the $\mathbb Z_2$ topological invariant of the honeycomb lattice in a manifestly gauge invariant way which allows us to include $S_z$ symmetry breaking terms -- like Rashba spin orbit interaction -- into the topological analysis. Interestingly, we find that the interplay of a Rashba and an intrinsic spin-orbit term can generate a non-trivial quantum spin Hall phase in silicene. This is in sharp contrast to the more extensively studied honeycomb system graphene where Rashba spin orbit interaction is known to compete with the quantum spin Hall effect in a detrimental way.
\end{abstract}

\maketitle

\section{Introduction}
One of the main subjects of current interest in condensed matter physics is the search for materials that host topological insulator (TI) phases \cite{HasanKane,XLReview2010,TSMReview}. Two dimensional TIs  exhibit the quantum spin Hall effect (QSHE) with gapless edge states and a finite energy gap in the bulk \cite{KaneMele,KaneMeleZ2,Bernevig}. The first proposal of this state of matter was made by Kane and Mele \cite{KaneMele} on the basis of graphene in the presence of spin-orbit interaction (SOI). However, the relevant SOI in graphene turns out to be rather small
\cite{Gmitra}
such that the effect seems to be inaccessible in experiments. This situation is different in HgTe/CdTe quantum wells where the QSHE was also predicted theoretically
\cite{Bernevig}
and experimentally seen soon after \cite{Koenig}.

Recently, a single layer of silicon atoms -- called silicene -- has been synthesized exhibiting an analogous honeycomb structure as graphene \cite{SilExp1,SilExp2,SilExp3}.  Since silicon is heavier than carbon, the spin-orbit gap in silicene is much larger than in graphene. Therefore, if it was possible at some point to prepare clean silicene, it should be feasible to experimentally access the QSHE in this material. Similar to graphene, the unit cell of silicene contains two atoms which gives rise to two different sublattices. In contrast to graphene, however, the silicene
sublattices are found to be arranged in a buckled structure pointing out-of-plane \cite{LiuSil}. Due to the broken sublattice symmetry, the mobile electrons in silicene are therefore able to couple differently to an external electric field than the ones in graphene. This difference is the origin of new (Rashba)
\footnote{We use the expression (Rashba) in brackets here to indicate, that some of the terms in question remind us of Rashba spin-orbit interaction terms, while others are of a different kind, like electric-field induced or intrinsic
spin-orbit terms.} 
spin-orbit coupling effects that allow for external tuning and closing of the band gap in silicene \cite{EzawaNjop}. Consequently, an electrically induced topological quantum phase transition is possible. It is natural to ask whether this phase transition can in principle go both ways, i.e., whether the electric field can be used to destroy and generate the QSHE. Refs. \cite{KaneMele,Ezawa2} clearly show that a different potential on the two sublattices of a honeycomb lattice leads to a transition from a TI to a trivial insulating state.  In Ref.~\cite{AnSil}, some indications have been presented that the interplay of two silicene specific (Rashba) spin-orbit terms 
can even induce the QSHE starting from a trivial insulating band structure in the absence of these terms.

The quantum spin Hall (QSH) phase is distinguished from a normal insulating phase by a  bulk $\mathbb Z_2$~topological invariant \cite{KaneMeleZ2,FuPump}. For a minimal model of the QSHE in graphene, this invariant has been analytically calculated in a seminal work by Kane and Mele \cite{KaneMeleZ2}. However, the original formulation of the $\mathbb Z_2$~ invariant in terms of Bloch functions does not contain a constructive prescription as to its numerical evaluation. Subsequent work on the topological properties of the band structure of silicene was restricted to the absence of terms breaking the spin $S_z$-conservation or to employing the bulk-boundary correspondence in silicene nanoribbons \cite{Ezawa2,AnSil}.

Evidently, a full topological analysis of silicene is missing and, as we show below, important to clearly identify phenomenological differences between graphene and silicene.
In this work, we employ Prodan's method \cite{Prodan} to calculate the topological invariant without any further symmetry assumptions in a manifestly gauge invariant way to provide a conclusive analysis of the novel features of silicene regarding QSH physics. In particular, we establish that, in contrast to graphene, the QSHE can be generated by (Rashba) SOI in silicene.

The bulk of this article consists of two parts which we keep fairly self contained to allow the reader to follow our analysis {\`a} la carte. In Sec.~2, we analyze the symmetries of the lattice of silicene. This analysis allows us to mathematically construct the low-energy Hamiltonian (close to the $K$ points of the Brillouin zone) by means of the invariant expansion method with a particular focus on terms involving a perpendicular electric field. Thereby, we discover for silicene an additional, electric field induced spin-orbit term of the low energy Hamiltonian. Furthermore, a tight-binding calculation is performed to verify the terms previously derived from symmetries and to estimate their magnitude. The reader who is more interested in quantum spin Hall physics can directly go to Sec.~3, where we study the topological properties of the band structure of silicene by explicitly calculating the $\mathbb Z_2$ topological invariant in a manifestly gauge invariant way. In this section, the possibility of a topological phase transition induced by an external field is carefully examined which enables us to correct previously proposed phase diagrams. Finally, we conclude in Sec.~4.
Some technical details of the invariant expansion and the tight-binding model are presented in the appendix.

\section{Symmetry based derivation of the Hamiltonian}
\subsection{Identification of the lattice symmetry}
\begin{figure}[htp]

\centering
 \includegraphics[width=0.8\linewidth]{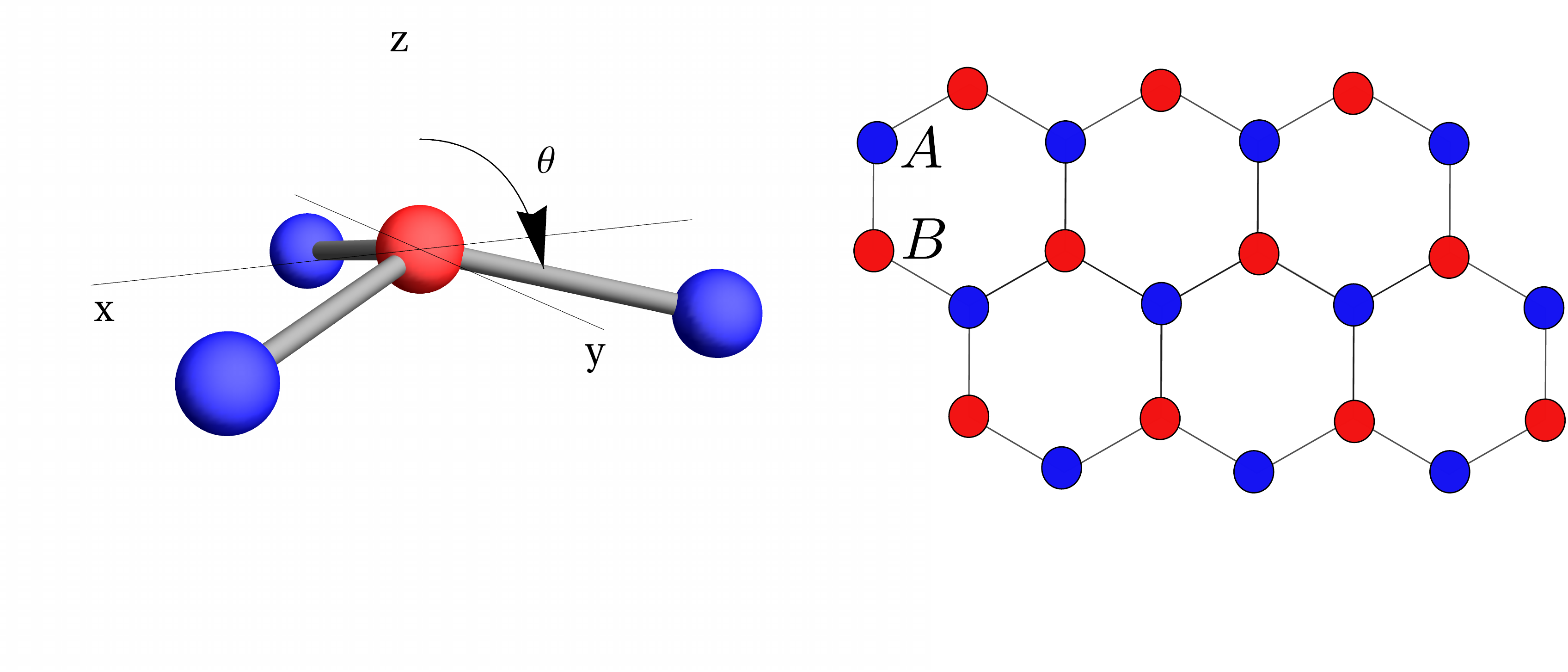}
\caption{\label{fig:silicene} Schematic of the real lattice of silicene. The sublattices (denoted $A$ and $B$) of the honeycomb structure are spatially separated in $z$-direction. The buckling-angle $\theta$ is found to be $101.7^{\circ}$ in a silicene lattice model \cite{LiuSil}.
}
\end{figure}

Silicene is a monolayer of silicon atoms arranged in a buckled honeycomb lattice (see Fig.~\ref{fig:silicene} for a schematic). In contrast to graphene, the two basis atoms of the unit cell
(called $A$ and $B$) are separated perpendicular to the atomic plane at a distance $2l$ with $l=0.23\mathring{A}$ \cite{Ezawa2}. As there is no translation symmetry in the out-of-plane direction, the material is quasi-two-dimensional. The buckling is quantified by an angle $\theta\geq90^{\circ}$ as shown in Fig.~\ref{fig:silicene}.

\begin{figure}[htp]
\centering
\includegraphics[width=.7\textwidth]{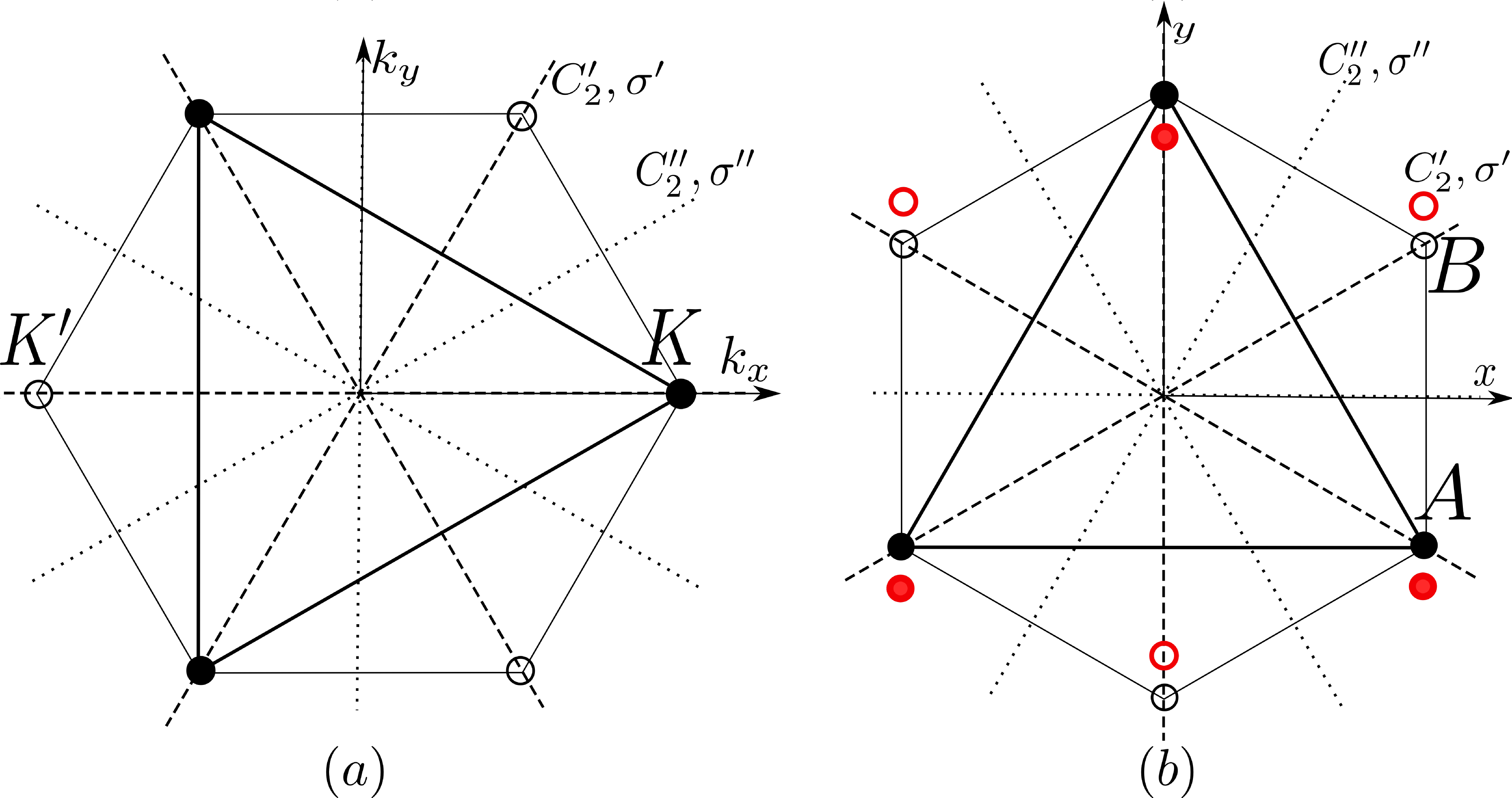}
\caption{\label{fig:geo} Symmetry operations, marked as dashed and dotted lines, of the reciprocal (a) and the real lattice (b). The operations $C_n$ denote rotations by $2\pi/n$ around the $z$-axis perpendicular to the plane; $C_n'$ and $C_n''$ refer to
rotations around the labeled axis, lying within the atomic plane. $\sigma$ describes reflection planes spanned by the labeled axis and the $z$-axis. The primed operations cross the corners of the underlying hexagon, while the double-primed ones do not. 
$A$ and $B$ denote different sublattices, $K$ and $K'$ inequivalent corner points
of the first Brillouin zone. The red dots implicate, that for silicene in real space the atomic sites are shifted perpendicular to the plane.}
\end{figure}

In Fig.~\ref{fig:geo}, we provide an illustration of the lattice of silicene -- in real and reciprocal space. 
Basis vectors defining the unit cell are given by
\begin{equation}\label{eq:myUnitVectors}
 \vec{a}_1=a \left(\frac{\sqrt{3}}{2},\frac{3}{2},0\right), \hspace{1cm} \vec{a}_2=a \left(-\frac{\sqrt{3}}{2},\frac{3}{2},0\right)
\end{equation}
in real space and by
\begin{equation}
 \vec{b}_1=\frac{2\pi}{a} \left(\frac{1}{\sqrt{3}},\frac{1}{3}\right), \hspace{1cm} \vec{b}_2=\frac{2\pi}{a} \left(-\frac{1}{\sqrt{3}},\frac{1}{3}\right)
\end{equation}
in reciprocal space. Here, $a$ is the distance between two neighboring silicon atoms.
%
In real space without buckling, the lattice exhibits $D_{6h}$ symmetry (i.e. the graphene case). All the symmetry operations sketched in Fig.~\ref{fig:geo} are present, as well as bulk inversion $i$ and reflection at the atomic plane itself, $\sigma_h$.
With the buckling present (silicene), symmetries $C_6$, $C_2$, $C_2'$, $\sigma''$ and $\sigma_h$ are broken and we are left with point group $D_{3d}$.\\
%
Let us now go to reciprocal space. For symmorphic groups, the $\Gamma$ point always has the same symmetry as the real lattice.
However, the notation in use of symmetry axes with a single prime referred to the fact, that these axes cross the corners of the underlying hexagon (see the caption of Fig.~\ref{fig:geo}). 
Under Fourier transformation, the hexagonal lattice is rotated by $90^{\circ}$ with respect to the axes of a fixed coordinate system. The positions of the $C_2$- and $\sigma$-axes stay the same then, while
the corners of the hexagon come to rest on the previously double-primed symmetry axes now. To have a corresponding group theoretical notation in reciprocal space as well, we rename the symmetry axes. Fourier transformation of the lattice can thus effectively be considered as an interchange of single and double-primed operations (see Fig.~\ref{fig:geo}).

%

In particular, the $D_{3d}$-symmetry at the $\Gamma$-point is equivalent to operations $C_3$, $C_2'$, $\sigma''$, $i$ and $S_6$ in reciprocal space, which corresponds to the point group $D_3$ with the additional symmetry classes $\{ \sigma'', i, S_6\}$. 
At the $K$ points of the Brillouin zone, the symmetry of the group of the wave vector is further reduced to the point group $D_3$.
%

%
\subsection{Invariant expansion}
The full knowledge of the lattice symmetries makes it in principle possible to construct a low-energy Hamiltonian by expansion around high-symmetry points. 
This can be done with powerful and well-established approaches, for instance, the invariant expansion \cite{PikusBir,Winkler,Roessler}. 
In undoped silicene, the Fermi level is located at the $K$-points of the Brillouin zone. Hence, a low-energy Hamiltonian constructed by a symmetry analysis near the $K$-points will capture the essential physical properties of the system.\\
We perform an invariant expansion around the $K$-points of silicene, which were identified to exhibit the symmetry point group $D_3$. The $\pi$-orbital wavefunction transforms like the two-dimensional IR $\Gamma_3$ of the group $D_3$, while
the spin part is represented by the IR $\Gamma_4$ of the double group. Therefore, our total wavefunction is of the form of the product $\Gamma_3 \times \Gamma_4^*=\Gamma_4+\Gamma_5$ and the starting point for 
the derivation of the Hamiltonian. A detailed presentation of this expansion is given in \ref{sec:AppA}.\\
Consequently, the low-energy Hamiltonian of silicene near the $K$-points is found to be (in lowest orders of $\vec{k}$ and $E_z$) 
\begin{eqnarray}\label{eq:Ham2}
\mathcal{H}^{K(K')} & = a_1\mathbb{I}+ a_2 \tau_z \sigma_z s_z  +  a
_3 \sigma_z s_0 E_z + a_4 \tau_z \sigma_0 s_z E_z + \\
& a_5 ( \tau_z \sigma_x s_y - \sigma_y s_x) E_z+  a_7 (\tau
_z \sigma_x k_x +\sigma_y k_y) s_0 + \nonumber \\
& a_{10} E_z [\sigma_x(s_x k_y + s_y k_x)+\tau_z\sigma_y(s_x k_x -s_y k_y)] \nonumber+\\
& a_{11} E_z \sigma_0 (s_x k_y - s_y k_x) + a_{13}\sigma_z (s_x k_y - s_y k_x) + \nonumber \\
& a_{16}E_z(\sigma_x k_x +\tau_z\sigma_y k_y)s_z, \nonumber
\end{eqnarray}
presented in the basis $(\psi_A \beta, \psi_A \alpha, \psi_B \beta, \psi_B \alpha)^T$, where $\alpha=|\uparrow\rangle$, $\beta=|\downarrow\rangle$. Here, $\tau_z=\pm 1$ distinguishes the inequivalent valleys $K$ and $K'$.
The Pauli matrices $\sigma$ act on sublattices $A$ and $B$, while $s$ are the corresponding matrices in spin space.
Additional constraints due to time-reversal symmetry (TRS) have 
already been taken into account, as explained in \ref{sec:AppA}.\\
In Eq.~(\ref{eq:Ham2}), the terms proportional to $a_2$, $a_5$, and $a_7$ are well known from the graphene literature to describe its fundamental properties near the $K$-points \cite{KaneMele}. Further corrections proportional to $a_{10}$ and $a_{11}$ can as well be found in graphene \cite{WinklerPa}. Beyond this, silicene exhibits specific terms proportional to $a_3$ and $a_{13}$ as reported in Refs. \cite{LiuSil, EzawaNjop}. 
Additionally, we find an electric field induced SOI-term proportional to $a_4$, that has not been reported for silicene before \footnote{Having completed our work on this manuscript, we became aware of a recent publication \cite{GmitraB}, that discusses a very similar term in the context of single-side semihydrogenated graphene.}.

Moreover, the Hamiltonian exhibits higher order
corrections to the linear dispersion (proportional to $a_{16}$). We conclude, that the low-energy Hamiltonian of silicene near the $K$-points contains interesting SOI terms that are absent in graphene because of the difference in the symmetry of the two honeycomb lattices.

\subsection{Corresponding tight-binding model}

To supplement the symmetry analysis, a corresponding tight-binding model is discussed next. This model enables us to construct a valid Hamiltonian for the full Brillouin zone which is crucial for the subsequent topological analysis.

Let us start with a brief discussion of the internal spin-orbit coupling and subsequently introduce the other important terms of the tight-binding model. In real space, silicene is described in terms of a lattice model including the (standard) spin-orbit coupling term \cite{KaneMele,LiuSil}
\begin{equation}\label{eq:HatomSo}
 H_{so}=\frac{\hbar}{4m_0c^2}(\vec{\nabla}V\times\vec{p})\cdot \vec{s}=-\frac{\hbar}{4m_0c^2}(\vec{F}\times\vec{p})\cdot \vec{s},
\end{equation}
where $\vec{F}$ is the force stemming from the electric potential $V$, $\vec{p}$ is the momentum, and $\vec{s}$ the spin of the electron. The (internal) electric force in silicene is provided by the crystal field in in-plane and out-of-plane direction. Since the momentum operator is oriented along nearest neighbor or next-nearest neighbor bonds, the silicene lattice forbids terms involving a crystal in-plane force coupling to a nearest neighbor bond. This consideration leads to the following spin-orbit Hamiltonian \cite{LiuSil}
\begin{eqnarray}\label{eq:Hso}
H_{so}&=
\imath \lambda_{so} \sum_{\langle\langle i,j\rangle\rangle; \alpha\beta} v_{ij} c_{i\alpha}^\dagger s_z^{\alpha\beta} c_{j\beta}-
\imath \frac{2}{3}\lambda_{r,2} \sum_{\langle\langle i,j\rangle\rangle; \alpha\beta} \mu_{i} c_{i\alpha}^\dagger (\vec{s}\times \hat{d}_{ij})_z^{\alpha\beta} c_{j\beta}
\end{eqnarray}
with, so far, undefined parameters $\lambda_{so}$ and $\lambda_{r,2}$. In the latter equation, $\alpha$ and $\beta$ are spin quantum numbers; the indexes $i,j$ label the atomic site/orbital. Here and in the following, $\langle i,j\rangle$ denotes nearest neighbors and $\langle\langle i,j\rangle\rangle$ next-nearest neighbors. The neighboring sites are each time connected by the vector
$\vec{d}_{ij}$ with its corresponding unit vector $\hat{d}_{ij}$. The sign $v_{ij}=\pm1$ refers to the next-nearest neighbor hopping being anticlockwise or clockwise with respect to the positive $z$-axis. Furthermore, $\mu_{i}=\pm1$ is introduced to distinguish between the $A(B)$ site.

Additionally, there is a regular nearest-neighbor hopping term and on-site energies of the form
\begin{equation}\label{eq:H0}
 H_0=\sum_{i\alpha} \epsilon_i c_{i\alpha}^\dagger c_{i\alpha}- \sum_{\langle i,j\rangle; \alpha\beta} t_{ij} c_{i\alpha}^\dagger c
_{j\beta},
\end{equation}
where $\epsilon_i$ is the on-site energy of the atomic orbital and $t_{ij}$ the hopping parameter for hopping between the orbitals $i$ and $j$ of
neighboring atomic sites. When an external electric field $E_z$ is applied perpendicularly to the atomic plane of silicene, a staggered sublattice potential of the form
\begin{equation}\label{eq:He}
 H_{E}=\lambda_e \sum_{i\alpha} E_z^i \mu_i c_{i\alpha}^\dagger c_{i\alpha}
\end{equation}
is generated \cite{Ezawa2} with a, so far, undefined parameter $\lambda_{e}$. Interestingly, $E_z^i$ allows for on-site transitions between the $p_z$ and $s$ orbitals \cite{Hill}.

To describe a full next-nearest neighbor tight-binding model, we also introduce spin-orbit terms involving external electric forces in Eq.~(\ref{eq:HatomSo}), which we may index as $H_R$ here due to their resemblance to Rashba terms. The simplest ones are
\begin{eqnarray}\label{eq:Hr}
H_R &=
\imath \lambda_{r,1} \sum_{\langle i,j\rangle; \alpha\beta} c_{i\alpha}^\dagger (\vec{s}\times\hat{d}_{ij})_z^{\alpha\beta} c_{j\beta} E_z^j +
  \imath \lambda_{e,2} \sum_{\langle\langle i,j\rangle\rangle; \alpha\beta} v_{ij} c_{i\alpha}^\dagger s_z^{\alpha\beta} c_{j\beta} \mu_{i} E_z^j+  \nonumber \\
&  \imath \lambda_{r,3} \sum_{\langle\langle i,j\rangle\rangle; \alpha\beta} c_{i\alpha}^\dagger (\vec{s}\times\hat{d}_{ij})_z^{\alpha\beta} c_{j\beta} E_z^j.
\end{eqnarray}
Interestingly, the second term proportional to $\lambda_{e,2}$ is a multiplicative combination of the spin-orbit term and
the staggered sublattice potential. This term corresponds to the additional, electric field induced spin-orbit term that we have found in the invariant expansion model (proportional to $a_4$).
The full tight-binding Hamiltonian is then given by
\begin{equation}\label{eq:FullLatticeHam}
 H=H_0+H_{so}+H_E+H_R.
\end{equation}
The terms given above will reproduce our Hamiltonian derived by symmetry analysis, Eq.~(\ref{eq:Ham2}), when being expanded around the $K$-points.

\subsection{Tight-binding model including $\pi$ and $\sigma$-bands of silicene}
To be able to estimate the coupling constants introduced above, we now briefly discuss and extend a tight-binding model presented in Ref.~\cite{LiuSil} including $\pi$ and $\sigma$-bands in silicene using the basis
\begin{equation} \label{eq:TbBasis1}
 \{| p_z^A \rangle,  |p_z^B \rangle, |p_y^A \rangle, |p_x^A \rangle, |s^A\rangle, |p_y^B \rangle, |p_x^B \rangle, |s^B\rangle \}\times\{\uparrow, \downarrow \}.
\end{equation}
Nearest-neighbor hopping matrix elements between given orbitals can then be expressed by
parameters $V_1$, $V_2$, and $V_3$ as functions of Slater bond parameters $V_{pp\pi}$, $V_{pp\sigma}$, $V_{sp\sigma}$ and curvature angle $\theta$ with
\begin{eqnarray*}
& V_1=\frac{3}{4} \sin^2 \theta (V_{pp\pi}-V_{pp\sigma}),\\
& V_2=\frac{3}{2} \sin\theta V_{sp\sigma}, \\
& V_3=\frac{3}{2} \sin\theta \cos\theta (V_{pp\pi}-V_{pp\sigma})
\end{eqnarray*}
(see Ref.~\cite{LiuSil} for details of the modeling).
Beyond the analysis done in Ref.~\cite{LiuSil}, we consider the influence of an electric field $E_z$ applied perpendicularly to the atomic plane.
A staggered sublattice-Hamiltonian is then induced that takes in the basis (\ref{eq:TbBasis1})
the following form
\begin{equation}
H_E= e E_z \left(
\begin{array}{cccccccc}
 0 & 0 & 0 & 0 & z_0 & 0 & 0 & 0 \\
 0 & 0 & 0 & 0 & 0 & 0 & 0 & z_0  \\
 0 & 0 & 0 & 0 & 0 & 0 & 0 & 0 \\
 0 & 0 & 0 & 0 & 0 & 0 & 0 & 0 \\
 z_0 & 0 & 0 & 0 & 0 & 0 & 0 & 0 \\
 0 & 0 & 0 & 0 & 0 & 0 & 0 & 0 \\
 0 & 0 & 0 & 0 & 0 & 0 & 0 & 0 \\
 0 & z_0  & 0 & 0 & 0 & 0 & 0 & 0
\end{array}
\right) \times \mathbb{I}_{(2\times2)},
\end{equation}
as only transitions between $p_z$ and $s$ orbitals of the same site are allowed \cite{Hill}. In this equation, $e$ is the
charge of an electron and $z_0$ the Stark element weighting the transition.
Spin-orbit coupling is now included in the same way as in Ref.~\cite{LiuSil} by the term $H_{SO}=\Delta \vec{L}\cdot\vec{s}$ where
$\vec{L}$ is the angular momentum vector, $\vec{s}$ the spin operator, and $\Delta$ the coupling constant.
The tight-binding model of $\pi$ and $\sigma$ bands allows us to estimate some of the parameters introduced in the previous section as a function of the parameters $V_{1-3}$ (that are in principle known) as well as the Stark element $z_0$.
To do so, we now apply a unitary transformation $U$ to the Hamiltonian that corresponds to the following change of basis
\begin{eqnarray}
\fl
& U^T  \{ p_z^A, p_z^B, p_y^A, p_x^A, s^A,p_y^B, p_x^B, s^B \}\times\{\uparrow, \downarrow \}=
\{ \phi_1, \phi_4, \phi_2, \phi_5, \phi_3, \phi_6, \phi_7, \phi_8 \}\times\{\uparrow, \downarrow \},
\end{eqnarray}
with mixed orbitals, for example, $\phi_1=u_{11} p_z^A + u_{21} s^A+ u_{31}(\frac{1}{\sqrt{2}}(p_x^B-\imath p_y^B))$.
Then, the transformed Hamiltonian
\begin{equation}
H'= U^\dagger (H_{SO}+H_E) U
\end{equation}
can be analyzed perturbatively on the first $(4\times4)$ block corresponding to the basis
$\{\phi_1 \uparrow, \phi_1 \downarrow, \phi_4 \uparrow, \phi_4 \uparrow \}$, which is expected to describe energy eigenstates near the Fermi energy \cite{LiuSil}.
Listing only terms including the external electric field, we find in first and second order perturbation theory the following three terms
\begin{eqnarray*}
H_{}^{(1,2)}=
\lambda_{e} \sigma_z s_0+ \lambda_{e,2} \sigma_0 s_z+\lambda_{r,1} (\sigma_x s_y + \sigma_y s_x)
\end{eqnarray*}
with parameters
\begin{eqnarray}
 &  \lambda_{e}= 2 e E_z z_0 u_{11} u_{21} \approx \frac{2 e E
_z z_0 V_3}{V_2}, \\
& \lambda_{e,2} \approx -\frac{\Delta  d e E_{z} z_{0} V_{3} }{2 V
_{2}^3}, \nonumber \\
& \lambda_{r,1} \approx \frac{\Delta  e E_{z} z_{0}}{2 V_{2}}. \nonumber
\end{eqnarray}
In the last step, the expressions were approximated under the assumption of low buckling, meaning $\theta\to 90^{\circ}$ and $V_3\to0$, where only
the term of lowest order in $V_3$ was kept.
The term proportional to $\lambda_{e}$ represents the on-site hopping occuring only in a buckled structure.
With the one proportional to $\lambda_{e,2}$, an electric field induced term of first order in SOI $\Delta$ is generated that is related to a next-nearest neighbor hopping. It depends linearly on
$V_3$, so this term is specific to the buckled silicene structure as well. Finally, the term proportional to $\lambda_{r,1}$ is the well-known Rashba spin orbit coupling reported already in Refs. \cite{KaneMele,KonschPa}. All terms given above agree nicely with our previous group theoretical analysis.
%
\subsection{Numerical estimates}
We now estimate the size of the spin-orbit terms coupling to an external electric field in silicene. The
bond parameters $V_{1-3}$, the energy $d$, the angle $\theta=101.7^{\circ}$, the lattice constant $a$, as well as the spin orbit interaction $\Delta$ are adapted from Ref.~\cite{LiuSil}.
We then approximate the Stark-element $z_0=\langle \phi_{n,0,0}|z|\phi_{n,1,0}\rangle$ as transition matrix elements between atomic wave functions, where in silicene we have $n=3$, since the outer shells are provided by the $3s$ and $3p$ orbitals. The corresponding wave functions were chosen as the wave functions of the hydrogen atom with the same quantum numbers.
We then find $z_0=3\sqrt{6}\times a_B/Z_{\rm eff}$ in silicene, where $a_B$ is the
Bohr radius and an outer electron experiences an effective atomic charge of $Z_{\rm eff}\approx 4.29$ \cite{ScreeningCore} due to screening. Hence, we estimate $z_0=0.906 \mathring{A}$.
For typical values of $E_z=50 eV/300 nm$, we calculate
\begin{eqnarray}
 & \lambda_e \approx 8.5\ meV, \\
 & \lambda_{e,2} \approx 12.8\ \mu eV, \nonumber \\
& \lambda_{r,1} \approx 22.7\ \mu eV. \nonumber
\end{eqnarray}
Thus, the additional term proportional to $\lambda_{e,2}$ is small compared to the term proportional to $\lambda_{e}$, but similar in magnitude as the Kane-Mele-Rashba term proportional to $\lambda_{r,1}$.
\section{Topological analysis}
Triggered by the theoretical prediction \cite{KaneMele,KaneMeleZ2,Bernevig} and experimental discovery \cite{Koenig} of the quantum spin Hall (QSH) effect, the study of topological effects in the physics of Bloch bands has been a major focus of condensed matter physics in recent years \cite{HasanKane,XLReview2010,TSMReview}. The QSH state is a bulk insulating state featuring metallic edge states that are protected by time reversal symmetry. Due to Kramers theorem, these edge states appear in so called helical pairs. The bulk energy gap in the original proposal for the QSH effect in graphene by Kane and Mele \cite{KaneMele} is due to an intrinsic SOI which preserves a residual $U(1)$~spin symmetry. In the presence of this global spin quantization axis, the helical edge states are characterized by a perfect locking of spin and momentum: States with opposite spin move with opposite chirality along the edge. In the presence of (Rashba) SOI, no spin symmetry is present resulting in the absence of a global spin quantization axis. Rather, the spin quantization axis of the edge states can precess spatially. However, as TRS is present the Kramers pair of helical edge states is still protected from hybridizing, i.e., from a gap opening on the edge. This statement is true as long as the bulk gap is maintained implying that the system is adiabatically connected to the $U(1)$~preserving case and hence is still in the QSH state. However, the role of Rashba SOI for the QSH effect in graphene is only detrimental: Upon increasing the strength of Rashba SOI, the system will go through a quantum phase transition destroying the QSH phase. Conversely, given the lattice symmetry of graphene, switching on Rashba spin orbit coupling cannot generate the QSH phase starting from a semi-metallic or trivial insulator phase.

The slightly reduced symmetry of silicene stemming from its buckled structure entails several new couplings, at least two of which are of key relevance regarding QSH physics. First, let us consider the staggered potential distinguishing sublattice A and B that has been introduced formally in Ref. \cite{KaneMele} to open a trivial gap. While this term is symmetry forbidden in graphene, it has been shown \cite{Ezawa2} to be induced by a simple out of plane electric field in silicene as we confirmed in our symmetry analysis above. This provides a knob to experimentally switch off the QSHE. Second, and even more interestingly, it has very recently been conjectured \cite{AnSil} that the QSHE can be generated by virtue of SOI terms that are symmetry forbidden in graphene but allowed in silicene. The authors of Ref. \cite{AnSil} probe the sub-gap conductance as a fingerprint of the quantum spin Hall state. While a non-vanishing conductance in a phase with a bulk gap is a promising signature of the QSH state, it is not in one to one correspondence with the $\mathbb Z_2$-invariant defining the quantum spin Hall effect \cite{KaneMeleZ2}. For example, what is called the QSHE2-phase in Ref. \cite{AnSil} is a trivial insulator which only shows the characteristic  conductance of $2\frac{e^2}{h}$~expected for the QSH phase since one additional pair of edge states does not contribute due to a mini-gap at the ribbon sizes considered in Ref. \cite{AnSil}. In the thermodynamic limit, the conductance in this parameter regime would be $4\frac{e^2}{h}$~signaling a $\mathbb Z_2$-trivial phase.

The main purpose of this section is to calculate the correct phase diagram of silicene in the presence of the (Rashba) SOI terms by rigorous calculation of the $\mathbb Z_2$~invariant in the absence of any symmetries besides TRS. We note that our method can be applied to study the entire parameter space of the silicene band structure without any further complications. However, in this work, we would like to focus on a particularly interesting parameter regime where the QSH phase is generated by (Rashba) SOI. Previous work on the topology of the QSH state in silicene in the presence of (Rashba) SOI \cite{Ezawa2} has been focused on effective models valid close to the $K$-points. However, in the presence of (Rashba) SOI, the bulk gap can close away from the $K$-points, a phase transition which would be missed by such power series expansions. Our analysis does not suffer from these limitations since we directly calculate the $\mathbb Z_2$-invariant characterizing the QSH phase from its very definition \cite{KaneMeleZ2,Prodan}. We find that there is indeed a QSH phase in the absence of the original Kane-Mele term $\lambda_{so}$. This QSH effect can be switched on by only tuning silicene specific $s_z$~symmtery breaking SOI terms and the staggered potential--all terms that are known to have only detrimental effect as to QSH physics in graphene. Our analysis, hence, settles in the affirmative the discussion on whether other SOI terms than the $s_z$~conserving Kane-Mele term can in principle generate a QSH state.


%
\subsection{Manifestly gauge invariant calculation of the topological $\mathbb Z_2$-invariant}
Let us briefly give an idea of how the topology of silicene may depend on an external electric field, orienting ourselves along the lines of Ref. \cite{Ezawa2}. We consider again the effective Hamiltonian of Eq.~(\ref{eq:Ham2}), derived by analytical expansion around the high-symmetry $K$-points. Without electric fields, the energy spectrum at $K$ exhibits a gap of size $|2 a_2|$.
Interestingly, in the presence of SOI, the bulk gap can be closed by an increasing electric field perpendicular to the plane. The gap closes approximately
at the very $K$-points, only if the corresponding parameter $\lambda_{r,1}$ is small compared to the transfer energy $t$, so $\lambda_{r,1}<<t$. The low-energy Hamiltonian allows to give an analytical expression
for the gap-closing critical field $E_z^c$ \cite{Ezawa2}.
At this point we may expect a topological phase transition from a QSH to a trivial insulating phase. This is indicated by the fact, that a gap is reestablished at the $K$-points, if the electric field is further increased, exceeding $E_z^c$.
However, a rigorous exploration of prospective topological quantum phase transitions in the silicene parameter space requires the calculation of a topological bulk invariant.

Therefore, we now turn to the general calculation of the topological $\mathbb Z_2$-invariant defining the QSH phase \cite{KaneMeleZ2}. While the $\mathbb Z_2$-invariant is of course a gauge invariant quantity by definition, the original literature \cite{KaneMeleZ2,FuPump} did not provide a constructive recipe for its numerical calculation. This is so because a calculation following the original definition requires a macroscopic gauge, though giving a gauge invariant result. By macroscopic gauge, we mean that the phase relation between Bloch functions at remote points in the Brillouin zone has to be fixed. However, when the band structure is calculated numerically, such phase relations are typically not accessible thus preventing the direct calculation of the invariant. This problem has only been resolved rather recently \cite{Prodan,Vanderbiltz2,XiaoLiangZ2} by a more constructive recipe for the direct calculation of the $\mathbb Z_2$-invariant. Here, we follow the method introduced by Prodan \cite{Prodan} which makes use of the elegant and manifestly gauge invariant formulation of the adiabatic connection \cite{TSMReview,Kato1950,Avron1988,AvronQuadrupole1989} originally introduced in a seminal work by Kato back in 1950 \cite{Kato1950}. Recently, the manifestly gauge invariant formulation of topological invariants has been generalized to other topological band structures \cite{TSMReview}. The crucial step of this approach consists in going from the Bloch functions of the occupied bands to the projection operator $P(k)$~onto the occupied states. As already pointed out by Kato in Ref. \cite{Kato1950}, the advantage of this construction is that the projection operator is obviously basis (gauge) independent.

The adiabatic connection is defined as
\begin{equation}
\mathcal A_\mu(k) = -\left[(\partial_\mu P(k)),P(k)\right],
\end{equation}
where $\partial_\mu=\frac{\partial}{\partial k^\mu}$~is the derivative in momentum space. Note that this connection is manifestly independent of the basis choice within the occupied bands in contrast to the more familiar Berry connection. Therefore, the adiabatic connection can be calculated numerically in a straight forward way which is in general not possible for the Berry connection. In Ref. \cite{Prodan}, the $\mathbb Z_2$~invariant of the QSH state is constructed in a similar way to Ref. \cite{FuPump} but using the adiabatic connection instead of the Berry connection. We refer the reader to Ref. \cite{Prodan} for this both accessible and fairly self contained explicit construction and review only the resulting expression for the $\mathbb Z_2$-invariant $\Xi$~for a rectangular lattice with lattice constants  $a_x,~a_y$, for completeness,
\begin{equation}
 \Xi= \alpha \frac{\det(U_{(0,0),(0,b_y)})\det(U_{(b_x, 0),(b_x, b_y)}) {\mathrm{Pf}}(\theta_{(0,0)}) {\mathrm{Pf}}(\theta_{(b_x,0)}) }
 {{\mathrm{Pf}}(\theta_{(0,b_y)}) {\mathrm{Pf}}(\theta_{(b_x,b_y)}) \sqrt{\det(U_{(0, b_y),(0,-b_y)}) }   \sqrt{\det(U_{(b_x, b_y),(b_x,-b_y)}) }  }.
\label{eqn:prodan}
\end{equation}
Here, $b_x=\frac{\pi}{a_x},~b_y=\frac{\pi}{a_y}$~are the boundaries of the Brillouin zone, $\theta_{(k_x,k_y)}$~is the matrix of the TRS operation which is anti-symmetric at the time-reversal invariant momenta ({TRIM}), and Pf denotes the Pfaffian. We note that $\Xi=-1$~defines the non-trivial QSH phase whereas $\Xi=1$~for a trivial insulator. The parameter $\alpha$~will be explained in detail below and
 \begin{equation}
 U_{(k^{(i)},k^{(f)})}=\mathcal T \mathrm{e}^{-\int_{k^{(i)}}^{k^{(f)}} \mathcal A}
 \label{eqn:KatoProp}
 \end{equation}
 describes the adiabatic evolution along the straight line from $k^{(i)}$~to $k^{(f)}$~with the path ordering operator $\mathcal T$. Unitaries generated by the adiabatic connection such as Eq.~(\ref{eqn:KatoProp}) can be conveniently calculated numerically as products of projection operators \cite{AvronQuadrupole1989,Prodan,TSMReview}. Explicitly,
\begin{equation}
 U_{(k^{(i)},k^{(f)})}=\lim_{n\rightarrow \infty}\prod_{j=0}^n P(k_j),\quad k_j= k^{(i)} + j~\frac{k^{(f)}-k^{(i)}}{n}. 
\label{eqn:ProjProd}
 \end{equation}
The path ordering appearing in Eq.~(\ref{eqn:KatoProp}) then amounts to the ordering of the product in Eq.~(\ref{eqn:ProjProd}) from the right to the left with increasing $j$. The construction resulting in Eq.~(\ref{eqn:ProjProd}) is similar to a Trotter decomposition for a time evolution operator and is correct to leading order in $\frac{1}{n}$. The gauge invariance of Eq.~(\ref{eqn:prodan}) might not be obvious at first glance due to the basis dependence of the Pfaffians. However, the combinations
\begin{equation}
\xi_{k_x}=\frac{\det(U_{(k_x,0),(k_x,b_y)}){\mathrm{Pf}}(\theta_{(k_x,0)})}{{\mathrm{Pf}}(\theta_{(k_x,b_y)})}=\pm \sqrt{\det(U_{(k_x, b_y),(k_x,-b_y)})}
\label{eqn:sqrtdet}
\end{equation}
with $k_x=0,b_x$~appearing in Eq.~(\ref{eqn:prodan}) are indeed gauge invariant up to the choice of the branch of the multivalued square-root as is obvious from the right hand side of Eq.~(\ref{eqn:sqrtdet}). That is the point where $\alpha$~comes into play. When calculating
\begin{equation*}
\Xi=\alpha\frac{\xi_{0}\xi_{b_x}}{\sqrt{\det(U_{(0, b_y),(0,-b
_y)})}\sqrt{\det(U_{(b_x, b_y),(b_x,-b_y)}}},
\end{equation*}
it has to be assured that the branch choices of the square-root at $k_x=0$~and $k_x=b_x$~are continuously connected to each other. This can be done by continuously interpolating the phase factor $\det(U_{(k_x, b_y),(k_x,-b_y)})$~between $k_x=0$~and $k_x=b_x$ \cite{Prodan}. If this phase has a winding such that the standard branch cut between Riemann sheets (line $(-\infty,0]$~in the complex plane) is crossed $n$~times during this interpolation, the naive result for $\Xi$~is corrected by the factor $\alpha=(-1)^n$.

While this procedure might be numerically challenging close to critical points or for large super-cells in disordered systems, it is basically a straightforward recipe. Note that the arbitrary basis choice for the representation matrix $\theta_{(k_x,k_y)}$~does not require any knowledge about the relative phase between the basis vectors at different points in $k$-space. Furthermore, the mentioned phase interpolation does not require any numerically inaccessible information
either, since the phase factor $\det(U_{(k_x, b_y),(k_x,-b_y)})$~at every $k_x$~is a gauge invariant quantity.

\subsection{From the honeycomb lattice of silicene to a rectangular super-cell}
\begin{figure}[htb]
\centering
\includegraphics[width=0.6\textwidth]{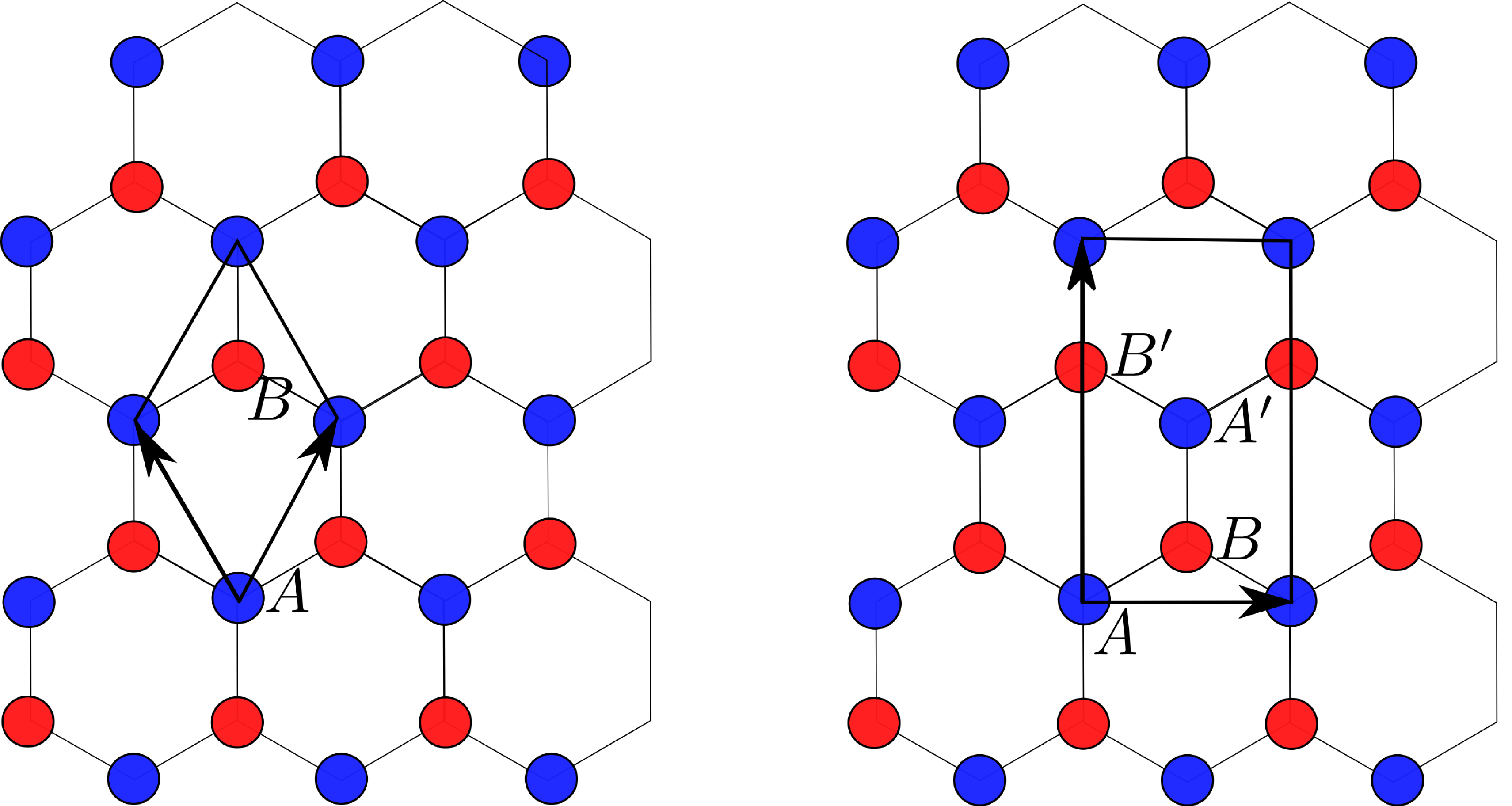}
\caption{\label{fig:unitC} Unit cells of the two-dimensional honeycomb lattice. On the left hand side, the minimal unit cell is drawn, containing two basis atoms $A$ and $B$.
On the right hand side, the unit cell has a rectangular form of doubled size. Here, four basis atoms $A$,$B$, $A'$, and $B'$ are included.}
\end{figure}

On the one hand, the procedure for the calculation of the $\mathbb Z_2$-invariant just described is general but requires a rectangular lattice whereas silicene crystalizes in a honeycomb lattice. On the other hand, the result for the topological invariant cannot depend on the choice of the unit cell. Therefore, we will now introduce a rectangular super cell which immediately allows us to apply the above method. To this end, the common basis vectors given in Eq.~(\ref{eq:myUnitVectors}) are combined to a new set of basis vectors
\begin{equation}
 \vec{a}'_1=\vec{a}_1+\vec{a}_2= \left(0,3 a,0\right), \hspace{1cm} \vec{a}'_2=\vec{a}_1-\vec{a}_2= \left(\sqrt{3} a,0,0\right),
\end{equation}
spanning a rectangular lattice with four atoms per site $A$, $B$, $A'$ and $B'$, as shown in Fig.~\ref{fig:unitC}. The Brioullin
zone is again rectangular with basis vectors
\begin{equation}
 \vec{b}'_1= \left(0,\frac{2 \pi}{3 a},0\right), \hspace{1cm} \vec{b}'
_2= \left(\frac{2 \pi}{\sqrt{3} a},0,0\right).
\end{equation}

The Bloch Hamiltonian of size $(8\times8)$~associated with our tight-binding model consists of the following contributions
\begin{equation}\label{eq:HTotal}
 H=H_t+H_{so}+H_e+H_{e,2}+H_{\lambda_{r,1}}+H_{\lambda_{r,2}}.
\end{equation}
The form of the $k$-dependent matrices is listed in Eqs.~(\ref{eq:AH1}) to (\ref{eq:AH6}) in \ref{sec:AppB}.
In the extended unit cell, the $K$-points (former corner points of the Brillouin zone) are mapped onto points $(\pm \frac{2 \pi}{3 \sqrt{3} a},0)$ inside the rectangle forming the new Brillouin zone.

For the Hamiltonian we chose the basis $\{|\psi_A\rangle, |\psi_B\rangle, |\psi_{A'}\rangle, |\psi_{B'}\rangle\} \times \{\uparrow, \downarrow \}$, where the
wavefunctions are of the form \cite{Bena}
\begin{equation*}
 |\psi_X \rangle =\frac{1}{\sqrt{N}}\sum_j e^{\imath \vec{k}\cdot \vec{R}_j} |\phi_j^X \rangle
\end{equation*}
with $X\in \{A,B,A',B'\}$. Here, $\phi$ are atomic wavefunctions, in particular $\langle \vec{r}|\phi_j^X\rangle= \phi (\vec{r}-\vec{R}_j^X)$.
$\vec{R}_j$ denote lattice vectors in real space, connecting the reference points of unit cells, while
$\vec{R}_j^X$ are the positions of atoms labeled $X$, in the unit cell $j$, relative to the reference point. The sum runs over all cells of the crystal.

\subsection{Results for the phase diagram of silicene}
Having constructed a rectangular lattice for silicene with four atoms per unit cell (see Fig.~\ref{fig:unitC}), we now investigate its topological phase diagram by direct evaluation of Eq.~(\ref{eqn:prodan}). The unitary adiabatic time evolutions (see Eq.~(\ref{eqn:KatoProp})) appearing in Eq.~(\ref{eqn:prodan}) are evaluated numerically using Eq.~(\ref{eqn:ProjProd}). 
The numerical calculations are done with a $k$-space discretization mesh of $n=200$~(see Eq.~(\ref{eqn:ProjProd})) and $200$~steps for the interpolation determining the phase winding $\alpha$~(see discussion below Eq.~(\ref{eqn:sqrtdet})). Close to the critical points, we have increased both discretizations from $200$~up to $1000$~steps to reach convergence. 
To ensure a regular evolution of the phase factor, we set the numerical threshold for the absolut value of $\det(U)$ to be greater than $0.7$.\\ 

\begin{figure}[htb]
\centering
\includegraphics[width=.60\textwidth]{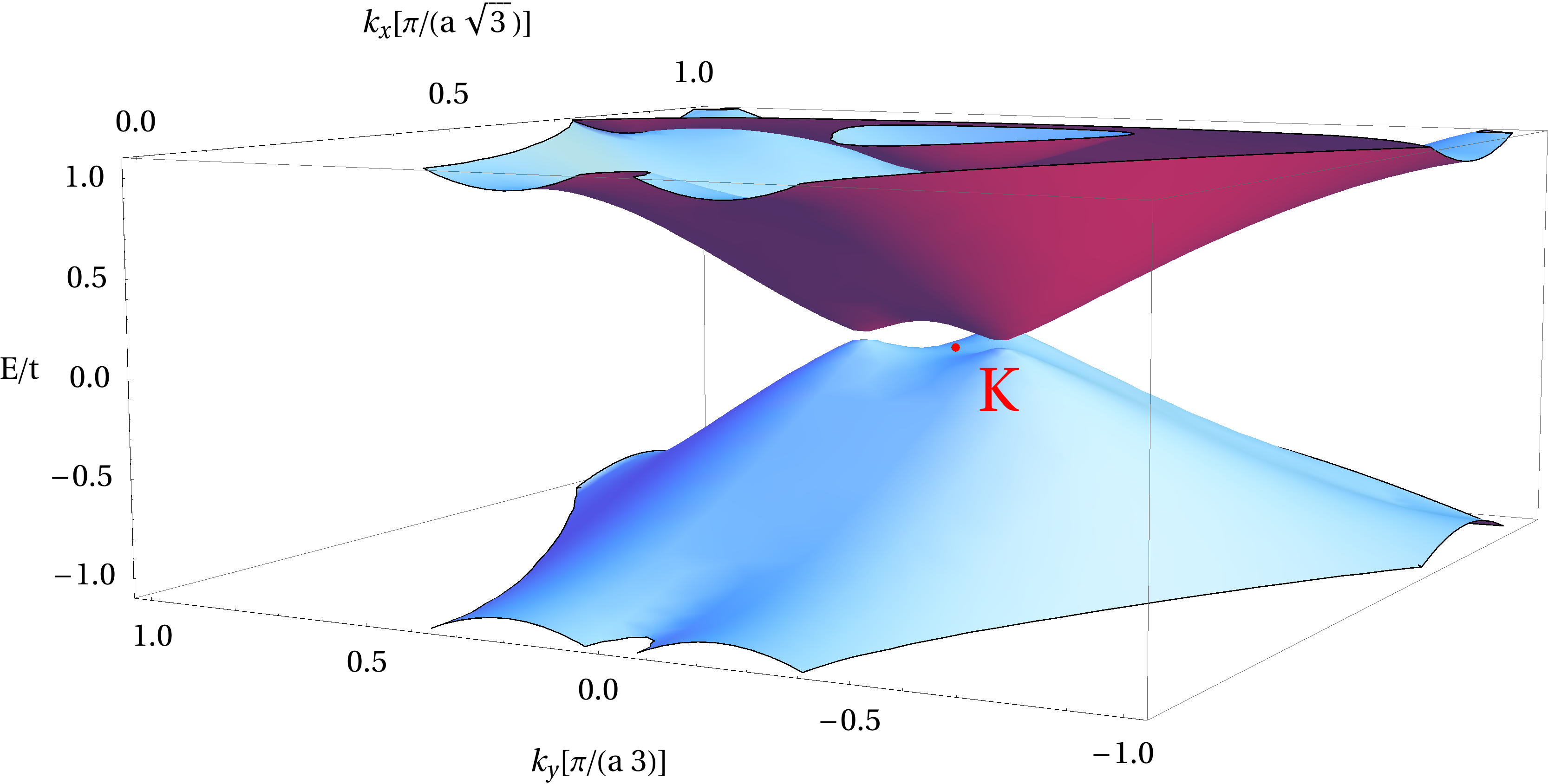}\\
\caption{\label{fig:ThreeMinGap} 
Plot of the bulk band structure of silicene with nonzero parameters $\lambda_e=0.1\ t$, $\lambda_{r,1}=0.3\ t $ and $\lambda_{r,2}=0.4\ t$. A rectangular unit cell is used, as explained in the text. Here, only half the Brioullin-zone 
is plotted due to implications of TRS. Under this parameter choice, the bulk band gap is about to close away from the $K$-points, which
correspond to the points $K(K')=(\pm\frac{2\pi}{3 \sqrt{3} a},0)$ (red dot).}
\end{figure}

\begin{figure}[htb]
\centering
\includegraphics[width=.60\textwidth]{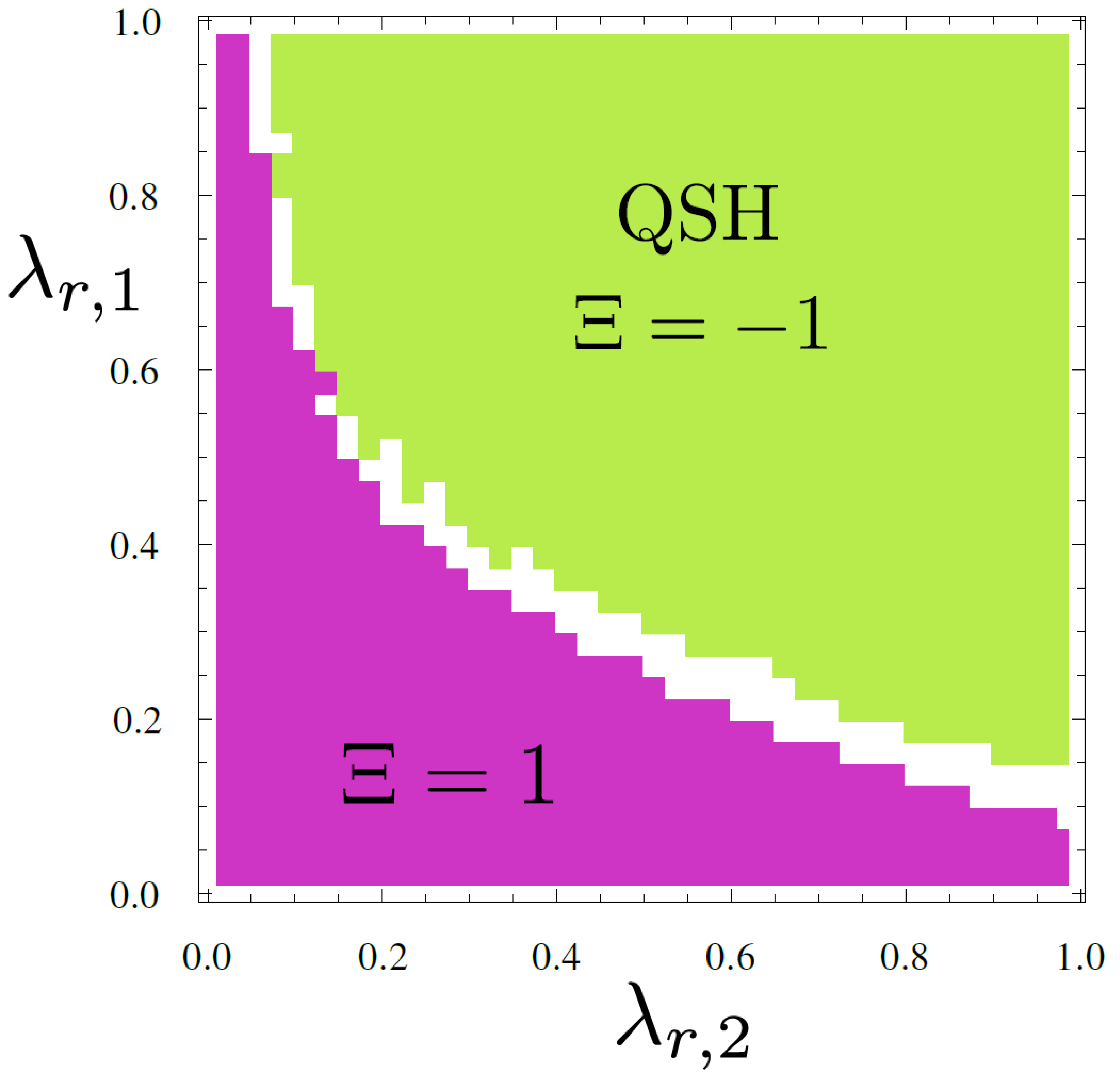}\\
\caption{\label{fig:Phases1} Topological phase diagram of silicene with free parameters $\lambda_{r,1},~\lambda_{r,2}$. The staggered potential $\lambda_e=0.1$~is fixed as well as $\lambda_{so}=0$. All couplings are measured in units of the hopping energy $t$. The white points close to the phase boundary denote critical regions where the bulk gap is so small that our numerical calculation did not converge properly.}
\end{figure}

\begin{figure}[htb]
\centering
\hspace{1.5cm} \begin{minipage}{.4\linewidth}

\includegraphics[width=1.1\textwidth]{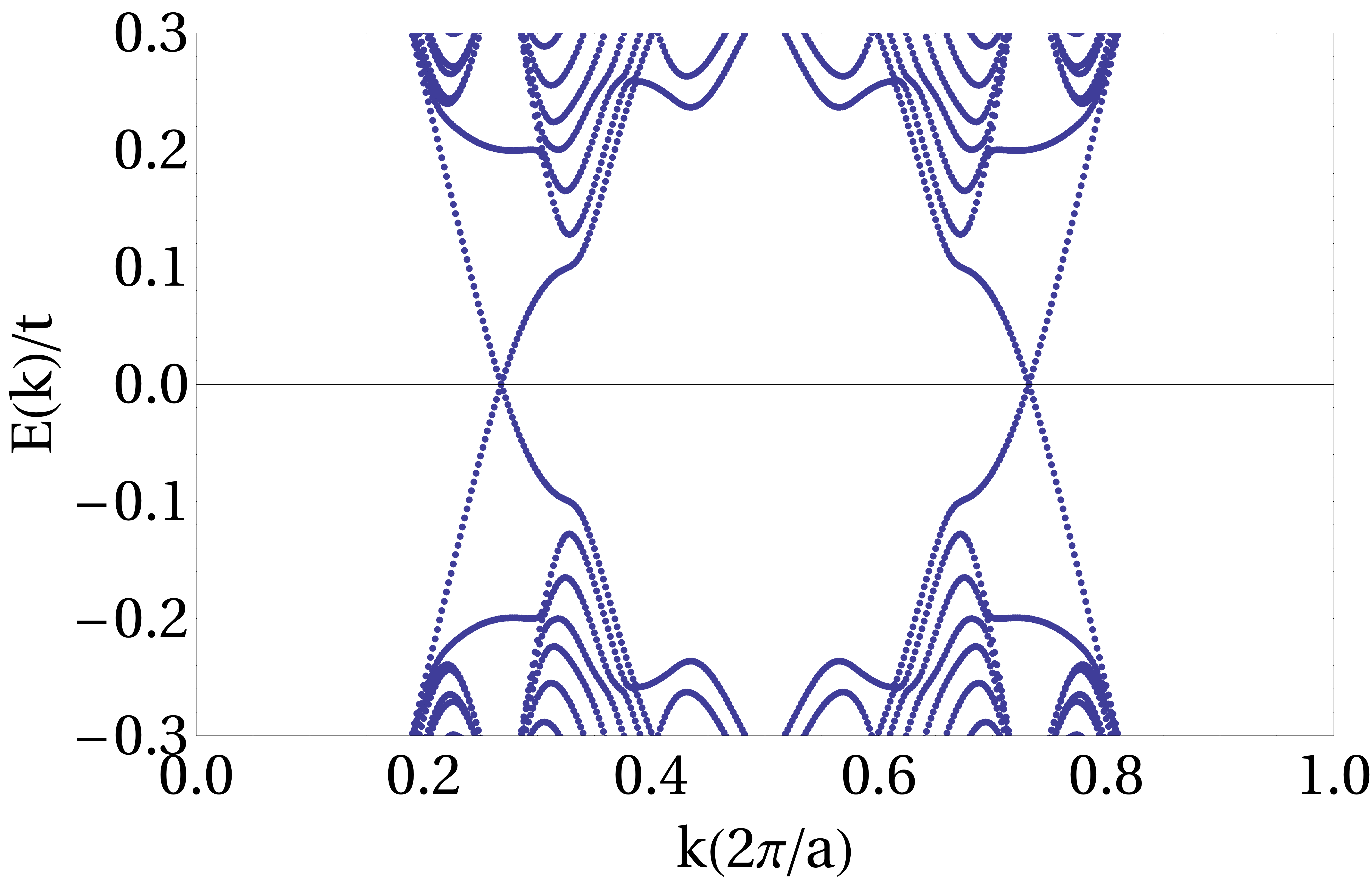}\\
\centering (a)
\end{minipage}
\begin{minipage}{.4\linewidth}

\includegraphics[width=1.1\textwidth]{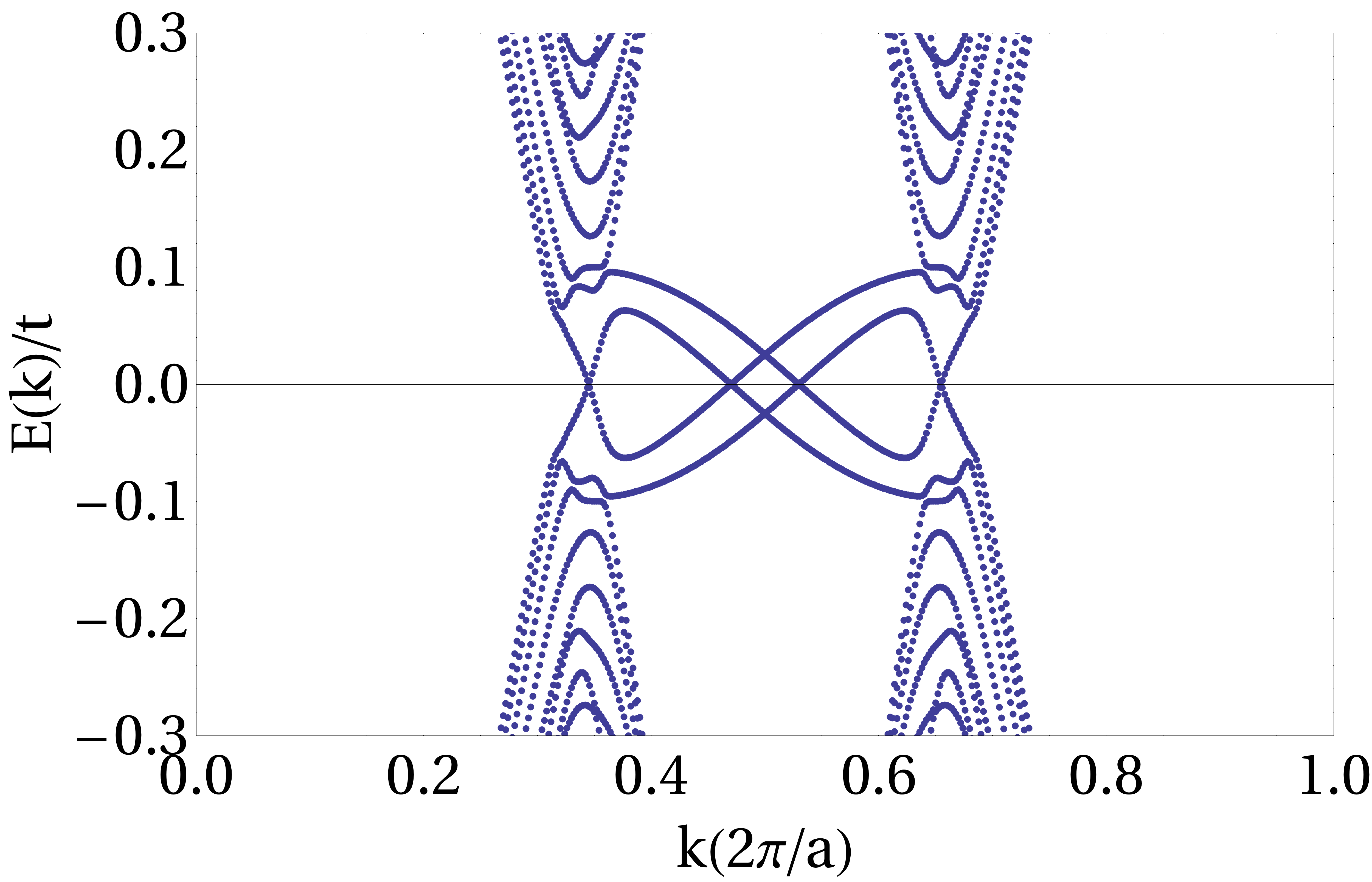}\\
\centering (b)
\end{minipage}

\caption{\label{fig:Qshe12} Simulation of the band dispersion of a silicene nanoribbon with zigzag-edges for two selected points of different regimes 
in parameter space. The unit cell corresponding to the width of the ribbon contains $N=104$ atoms. The numerical calculation was based on a discretization of the $k$-space in $500$ steps. 
(a) $\lambda_e=0.1\ t$, $\lambda_{r,1}=1.0\ t $ and $\lambda_{r,2}=0.2\ t$.
(b) $\lambda_e=0.1\ t$, $\lambda_{r,1}=0.05\ t $ and $\lambda_{r,2}=0.6\ t$. In the QSH-regime (a), there is an odd number of pairs of edge states at each edge. On the other hand, an additional pair of edge states leading to an even number of pairs, causes the topology to be trivial 
in the regime represented by (b).} 
\end{figure}

For notational simplicity, the electric field $E_z$ is absorbed in the parameters $\lambda_i$, that is $\lambda_i E_z\to \lambda_i$. All parameters are measured in units of the hopping matrix element $t$, which is known to be of the order of $1.6\ eV$ in silicene \cite{EzawaNjop}. We keep in mind that $\lambda_e$, $\lambda_{e,2}$, and $\lambda_{r,1}$ depend on the external electric field. We first benchmark the general method in a parameter regime where we have a good intuition for the expected QSH physics from previous literature \cite{KaneMele,Ezawa2}. Let us start with just one non-vanishing parameter $t$, all other parameters
being zero. This choice corresponds to the well known gapless Dirac cones at the $K$-points. Upon switching on $\lambda_{so}$ \cite{KaneMele}, the system exhibits a bulk gap of size $|2\lambda_{so}|$ and we reproduce $\Xi=-1$~\cite{KaneMeleZ2}.
We now add a term proportional to $\lambda_e$~depending on the external field. Upon increasing $\lambda_e$~the bulk gap
closes at the $K$-points for a critical value as predicted before. For even stronger fields a trivial gap opens, i.e., $\Xi=+1$.
Similarly, the gap closes for terms proportional to $\lambda_{e,2}$. Next, we add the Rashba-term proportional to $\lambda_{r,1}$ and the intrinsic SOI-term proportional to $\lambda_{r,2}$.
The first term $\lambda_{r,1}$ tends to establish three additional minima of the energy gap,
that are shifted away from the $K$-point. On the other hand, $\lambda_{r,2}$, primarily providing gapless states at the corners of the Brillouin zone (BZ), causes a modification of the band slope.
From an analysis of the overall band structure, one finds, that the interplay of both terms $\lambda_{r,1}$ and $\lambda_{r,2}$ may possibly close the band gap away from the $K$-points (see Fig.~\ref{fig:ThreeMinGap}).\\
After these general considerations, we would now like to focus on the interesting parameter regime identified in Ref. \cite{AnSil}, i.e., we consider $\lambda_e,~\lambda_{r,1},~ \lambda_{r,2}$~as free parameters that are measured in units of $t$. All other parameters are set to zero. Unfortunately, the analysis in Ref.~\cite{AnSil} was not suitable to predict the correct phase diagram for the $\mathbb Z_2$~invariant. However, our more rigorous analysis confirms the general conjecture that in this regime, a combination of $\lambda_{r,1}$~and $\lambda_{r,2}$~is able to drive a topological quantum phase transition away from the $K$-points which induces a non-trivial QSH phase.
If we only switch on $\lambda_e$, a trivial gap of size $|2\lambda_{e}|$~ opens. By tuning $\lambda_{r,1}$~and $\lambda_{r,2}$, the bulk gap can be closed away from the $K$-points, and a gap characterized by $\Xi=-1$, i.e., a QSH state emerges. We present the phase diagram in the identical parameter regime as presented in Ref.~\cite{AnSil} in Fig.~\ref{fig:Phases1}. Our direct calculation of the $\mathbb Z_2$-invariant disagrees with Ref. \cite{AnSil} both qualitatively (absence of the QSHE2) and quantitatively (significantly different phase boundary for the QSH phase). However, we would again like to point out that we can definitely confirm the phenomenology of a QSH phase in the absence of the Kane-Mele term $\lambda_{so}$. Instead, the extended QSH region shown in Fig.~\ref{fig:Phases1} is only driven by the two silicene specific (Rashba) SOI terms $\lambda_{r,1}$~and $\lambda_{r,2}$~which is conceptually 
very interesting. The existence of gapless edge states was additionally verified by the simulation of zigzag-edge silicene nanoribbons, as shown in Fig.~\ref{fig:Qshe12}. However, we emphasize, that our topological analysis is 
based on bulk properties and does not depend on specific forms of the boundaries.\\
Lastly, we found the phase diagram in Fig.~\ref{fig:Phases1} to be robust against small perturbations proportional to the term $\lambda_{so}$. For larger Kane-Mele parameters $\lambda_{so} > \lambda_e/(3 \sqrt{3})$, another non-trivial regime $(\Xi=-1)$ enters the phase diagram. This regime occurs independently of $\lambda_{r,2}$ and can be expelled again by increasing $\lambda_{r,1}$. Essentially, the
previously described quantum phase transitions are not affected.\\

\section{Conclusion}
A symmetry analysis of the lattice of silicene close to the $K$ points of the Brillouin zone was performed. With the help of the invariant expansion model, the $\pi$-band Hamiltonian
was constructed by symmetry considerations only, including spin-orbit coupling and external electric fields perpendicular to the atomic plane. Thereby, we discovered an additional term that is allowed by symmetry and related to an interplay of spin-orbit coupling and an external electric field in a buckled honeycomb structure. We supplemented the symmetry analysis by a tight-binding model which allowed us to estimate the relevant parameters of the model.

Subsequently, we carefully analyzed the topological properties of the band structure and proved that a topological phase transition can be generated by an external electric field. This analysis enabled us to plot a topological phase diagram of silicene employing Prodan's manifestly gauge-invariant method for the direct calculation of the $\mathbb Z_2$~topological invariant defining the quantum spin Hall phase. Since this method requires a rectangular lattice, we considered a rectangular silicene super-cell that contains four atoms.
Interestingly, the tunable phase transition can happen in a destructive as well as a constructive way, i.e., a quantum spin Hall phase can not only be destroyed but also be generated by means of external (Rashba) spin orbit coupling.
\section*{Acknowledgments}
We have benefitted from discussions with Roland Winkler. This work has been financially supported by the Deutsche Froschungsgemeinschaft (DFG-JST Research Unit ``Topotronics'', Priority Program ``Topological Insulators''), the European Science Foundation, the Helmholtz Foundation (VITI), and the Swedish Research Council.

\clearpage
\appendix

\section{Invariant expansion}
\label{sec:AppA}

\subsection{General theory}
%
Let us, for completeness, shortly review the general theory of the invariant expansion.

In this approach, the Hamiltonian is constructed to be invariant under
all symmetry operations of the point group of the underlying lattice. When spin is included, it is composed block-wise according to irreducible representations (IR) of the corresponding double group.
Each block $\mathcal{H}^{\alpha \beta}(\vec{\mathcal{K}})$, that may depend on some general tensor components $\vec{\mathcal{K}}$, has to
reflect the symmetry properties of all the IRs $\Gamma_{\kappa}$, that are contained in the product
\begin{equation*}
 \Gamma_{\alpha}\times \Gamma_{\beta}^*=\sum_{\kappa} n_{\kappa} \Gamma_{\kappa},
\end{equation*}
where the integer $n_{\kappa}$ is the multiplicity. The most general ansatz is then
\begin{equation} \label{eq:SymMatr}
 \mathcal{H}^{\alpha \beta}(\vec{\mathcal{K}})=\sum_{\kappa, \lambda, \mu} a_{\lambda\mu}^{\alpha\beta, \kappa} \sum_{l} X_{l}^{(\kappa, \lambda)} \mathcal{K}_l^{(\kappa, \mu)\ *}.
\end{equation}
Here, $\kappa$ runs over all IR $\Gamma_{\kappa}$ included in the product and $a_{\lambda\mu}^{\alpha\beta, \kappa}$ are material-specific constants.
 The symmetrized tensor operator components $\mathcal{K}_l^{(\kappa, \mu)\ *}$ are parameters like the wave vector
$\vec{k}$, as well as the external magnetic $\vec{\mathcal{B}}$ or electric field $\vec{\mathcal{E}}$. Furthermore, $X^{(\kappa, \lambda)}_l$ are the symmetrized matrices we wish to find to construct the Hamiltonian. Evidently, $\lambda$ und $\mu$ are indices numbering the different possible matrix- and
tensor-components. These indices run from unity to limits that depend on the order of the expansion. One set of the $\kappa$, $\lambda$, and $\mu$ is called an invariant.

The irreducible tensor components belonging to an IR are composed analogously to its eigenfunctions.
With the help of projection operators \cite{Dresselgroup}, that contain the matrix representations of each IR, we can combine tensor components at any order
and project them onto the appropriate IR of the point group. To construct the basis matrices $X^{(\kappa, \lambda)}_l$, we can use the Wigner-Eckart theorem \cite{Sakurai}. Since any point group is a subgroup of the full rotation group $\mathcal{R}$, the eigenfunctions of any IR of the point group are also eigenfunctions of $\mathcal{R}$ and
can be written in terms of spherical harmonics. Angular momentum quantum numbers are assigned to each IR, by defining the axial vector components $R_x$, $R_y$, and $R_z$ to be
the real angular momentum eigenstates with quantum number $l=1$.
Each IR can now be classified with angular momentum quantum numbers by comparing its eigenfunctions to a table of real spherical harmonics.
A symmetrized matrix $X^{k\ *}_{\nu}$ (that transforms like the IR $\Gamma_{k}^*$ of dimension $l_{k}$
and numbering $\nu=1,\ldots l_{k}$) which is contained in the block $\Gamma_i^* \times \Gamma_j$ is then derived
by Clebsch-Gordan coefficients (CGC)
\begin{equation}\label{eq:ccbasmatr}
 (X^{k *}_{\nu})_{\lambda \mu}=\left(
\begin{array}{cc|c}
k & j & i\ l \\
\nu & \mu & \lambda \\
\end{array}
\right)^*,
\end{equation}
where the order of $i$, $j$, and $k$ is crucial. In Eq.~(\ref{eq:ccbasmatr}), $l$ is again the multiplicity, that shows how often the IR $\Gamma_{k}$ is included in the product $\Gamma_i^* \times \Gamma_j$. For multiplicities
larger than one, we find linear independent sets of basis matrices.
%
\subsection{Application to the $\pi$-band Hamiltonian of silicene near the $K$ points}
We now start to perform an invariant expansion of the group $D_3$, that we found to be the group of the wave vector at the $K$-points in silicene (see Ref.~\cite{Koster} for a general discussion of this point group).
We are interested in the two-dimensional $\pi$-bands of silicene. The orbital part of the wavefunction corresponds to the two-dimensional IR $\Gamma_3$ of the group $D_3$. 
Since we wish to include spin in our symmetry analysis, we have to complement this IR by $\Gamma_4$, the appropriate double group IR of $D_3$. Consequently, we start from the following product of representations $\Gamma_3 \times \Gamma_4^*=\Gamma_4+\Gamma
_5.$ The resulting $(4\times4)$ Hamiltonian based on the invariant expansion can then be composed in blocks of four $(2\times2)$ matrices
\begin{equation}\label{eq:blocksD3}
\mathcal{H}=
\left(
\begin{array}{cc}
 \mathcal{H}_{44} &  \mathcal{H}_{45} \\
 \mathcal{H}_{54} & \mathcal{H}_{55}
\end{array}
\right).
\end{equation}
The construction by CGCs (as discussed in the previous section) implies that this $\pi$-band Hamiltonian is given in the basis $(\psi_A \beta, \psi_B \alpha, -\psi_A \alpha, \psi_B \beta)^T$, with $\psi_{A(B)}\equiv R_x\mp\imath R_y$ and $\alpha=|\uparrow\rangle$, $\beta=|\downarrow\rangle$. Each block of the Hamiltonian is thus decomposed into IRs.
General tensor components of any desired order can be found with the help of projection operators from the point group character table.
In Table~\ref{tab:expIrrCompD3}, we list components of interest up to third order.
\begin{table}[h]
\caption{\label{tab:expIrrCompD3} Character table and irreducible tensor components of $D_{3}$ up to third order in the tensor components $\vec{x}$. $x,y,z$ denote polar vector components and $R_x, R_y, R_z$ axial vector components. Here, axial and polar vector components appear on equal footing as the symmetry operations of $D_3$ do not provide mirror planes. Therefore, for simplicity, combinations with components of $\vec{R}$ are not listed again in higher orders but they are of course present.}
\begin{indented}
\lineup
\item[] \begin{tabular}{@{} llllp{2.4cm}p{4cm}}
\br
 & $E$ & $2C_3$ & $3C_2'$ & & \\
\mr

$\Gamma_1$ & $1$ & $1$ & $1$ &  & $x^2+y^2$; $z^2$  \\
$\Gamma_2$ & $1$ & $1$ & $\-1$ & $R_z$; $z$ &  \\
$\Gamma_3$ & $2$ & $\-1$ & $0$ & $(x,y)$; $(R_x,R_y)$ & $(yz,-xz)$;$(y^2-x^2,xy+yx)$ \\
\br
\end{tabular}\\
\item[] \begin{tabular}{@{} llllp{6.85cm}}
\br
 & $E$ & $2C_3$ & $3C_2'$ & \\
\mr
$\Gamma_1$ & & & & $z(x^2+y^2)$; $x(3y^2-x^2)$ \\
$\Gamma_2$ & & & & $y(y^2-3x^2)$ \\
$\Gamma_3$ & & & & $(x(x^2+y^2),y(x^2+y^2))$;
$((y^2-x^2) z,2xyz)$ \\
\br
\end{tabular}
\end{indented}
\end{table}
Identifying combinations of general tensor components of $\vec{R}$ with angular momentum quantum numbers, we can assign such quantum numbers to the
IRs itself. The basis matrices are then identified from CGCs according to Eq.~(\ref{eq:ccbasmatr}) and listed in Table~\ref{tab:expBasMatrD3}. Note that we have applied a unitary basis transformation
to guarantee a more symmetric basis  $(\psi_A \beta, \psi_B \alpha, \psi_A \alpha, \psi_B \beta)^T$. This transformation is done for comparison with the graphene case carefully analyzed in Ref.~\cite{WinklerPa}.
\begin{table}[h]
\caption{\label{tab:expBasMatrD3} Symmetrized matrices for blocks arising from the product $\Gamma_3\times \Gamma_4^*=\Gamma_4+\Gamma_5$ of the double group of $D_{3}$. Our choice of basis is
$(\psi_A \beta, \psi_B \alpha, \psi_A \alpha, \psi_B \beta)^T$. $\sigma_i$ denote the $(2\times2)$ Pauli matrices with $i \in \{x,y,z\}$.}
\begin{indented}
\item[] \begin{tabular}{@{} ll p{4cm}}
\br
$\mathcal{H}_{44}$ & $\Gamma_4\times\Gamma_4^*=\Gamma_1+\Gamma_2+\Gamma_3$ & $ \Gamma_1: \mathbb{I}$ \newline
$\Gamma_2: \sigma_z$ \newline
$\Gamma_3: (\sigma_x, -\sigma_y)$\\
$\mathcal{H}_{55}$ & $\Gamma_5\times\Gamma_5^*=2\Gamma_1+2 \Gamma
_2$ & $ \Gamma_1: \mathbb{I}; \sigma_x$ \newline
$\Gamma_2: \sigma_z; \sigma_y$ \\
$\mathcal{H}_{45}$ & $\Gamma_4\times\Gamma_5^*=2\Gamma_3$ & $ \Gamma_3: (\mathbb{I},-\imath \sigma_z);(\sigma_x, \sigma_y)$ \\
\br
\end{tabular}
\end{indented}
\end{table}
We use the symmetrized matrices and tensor components listed in Tables~\ref{tab:expIrrCompD3} and \ref{tab:expBasMatrD3} to
expand the Hamiltonian up to first orders in $\vec{k}$ and the electric field $E_z$ perpendicular to the plane. 
Note that since $\mathcal{H}$ is Hermitian, the
coefficients of the diagonal blocks have to be real, while those of the off-diagonal blocks can in principle be imaginary. We write $\gamma$ and $\imath \gamma'=\gamma$ to
include both cases with real coefficients $\gamma$, $\gamma'$. Then, we obtain for the Hamiltonian
\begin{eqnarray}\label{eq:H44grC3v}
\mathcal{H}_{44}= &\alpha_1 \mathbb{I} + \alpha_2 \sigma_z E_z + \alpha_3 (\sigma_x k_x - \sigma_y k_y)+\alpha_4 E_z (\sigma
_x k_y +\sigma_y k_x),
\end{eqnarray}
\begin{eqnarray}\label{eq:H55grC3v}
\mathcal{H}_{55}= &\beta_1 \mathbb{I}  + \beta_2 \sigma_x 1 +\beta_3 \sigma_z E_z + \beta_4 \sigma_y E_z,
\end{eqnarray}
\begin{eqnarray} \label{eq:H45grC3v}
 \mathcal{H}_{45}=& \gamma_1 (\mathbb{I} k_x -\imath \sigma_z k_y)+\imath \gamma_1' (\mathbb{I} k_x -\imath \sigma_z k_y) + \\
&  \gamma_2 (\sigma_x k_x + \sigma_y k_y)+
 \imath \gamma_2' (\sigma_x k_x + \sigma_y k_y)+ \nonumber \\
& \gamma_3 E_z (\mathbb{I} k_y +\imath \sigma_z k_x)+ \imath \gamma_3' E_z (\mathbb{I} k_y +\imath \sigma_z k_x)+ \nonumber \\
& \gamma_4 E_z (\sigma_x k_y -\sigma_y k_x) +
\imath \gamma_4' E_z (\sigma_x k_y -\sigma_y k_x). \nonumber
\end{eqnarray}
For a better physical interpretation, we once more change the basis by an unitary transformation and present the total Hamiltonian in the basis
$(\psi_A \beta, \psi_A \alpha, \psi_B \beta, \psi_B \alpha)^T$.
After this basis transformation, the Pauli matrices $\sigma$ and $s$ obtain the following physical meaning: $\sigma$ acts on the space of sublattices $A$ and $B$, and $s$ on the spin space.
In lowest order, the Hamiltonian in our new basis exhibits sixteen terms labeled by coefficients $a_1$ to $a_{16}$
\begin{eqnarray}\label{eq:Ham}
\mathcal{H}^K & = a_1\mathbb{I}+ a_2  \sigma_z s_z  +  a_3 \sigma_z s_0 E_z +
a_4  \sigma_0 s_z E_z + a_5 (  \sigma_x s_y - \sigma_y s_x) E_z+  \\
& a_{6} ( \sigma_x s_x + \sigma_y s_y) +a_7 ( \sigma_x k_x +\sigma_y k_y) s_0 + a_8  \sigma_0 (s_x k_x +s_y k_y)+ \nonumber \\
& a_9[\sigma_x(s_x k_x - s_y k_y)-\sigma_y(s_y k_x +s_x k_y)] + \nonumber \\
& a_{10} E_z [\sigma_x(s_x k_y + s_y k_x)+\sigma_y(s_x k
_x -s_y k_y)] +\nonumber \\
& a_{11} E_z \sigma_0 (s_x k_y - s_y k_x) + a_{12} E_z (\sigma_x k_y - \sigma_y k_x) s_0 + \nonumber \\
& a_{13}\sigma_z (s_x k_y - s_y k_x) +  a_{14} (\sigma_x k
_y - \sigma_y k_x) s_z + \nonumber \\
& a_{15} E_z \sigma_z (s_x k_x+s_y k_y)+a_{16}E_z(\sigma
_x k_x +\sigma_y k_y)s_z. \nonumber
\end{eqnarray}
Note, that the coefficients in Eqs.~(\ref{eq:H44grC3v}) to Eq.~(\ref{eq:H45grC3v}) and in Eq.~(\ref{eq:Ham}) are consistent, in the sense, that they are related by mutual linear combinations.
Here, we would already like to point out the additional, interesting spin-orbit term proportional to the coefficient $a_4$, coupling directly the out-of-plane components of spin and electric field.

\subsection{Consequences of time-reversal symmetry}
The terms derived so far in the invariant expansion have not yet been checked for consistency with time reversal symmetry which holds throughout this work since we only consider the influence of electric fields. Importantly, the point group $D_3+\sigma''$ provides symmetry operations that map the inequivalent corner points $K$ and $K'$ of the Brillouin zone onto each other. For example,
we can consider the reflection at a plane perpendicular to the atomic plane and including the y-axis (called $R_y$ in Ref.~\cite{WinklerPa}). This symmetry operation has the following impact on the Hamiltonian
\begin{equation*}
 \mathcal{D}(R_y) \mathcal{H}^K(\vec{\mathcal{K}}) \mathcal{D}(R_y)^{-1}=\mathcal{H}^{K'}(R_y^{-1}\vec{\mathcal{K}}),
\end{equation*}
where $\mathcal{D}(R_y)$ is the matrix representation of the operation $R
_y$. Polar ($\vec{x}$) and axial ($\vec{R}$) tensor components will then transform like
$R_y^{-1}: (x,y,z)=(-x,y,z)$ and $R_y^{-1}: ({R}_x,{R}_y,{R}_z)
=({R}_x,-{R}_y,-{R}_z)$. The
sublattices will not be changed by $R_y$. In a compact notation, we can write $R_y=\sigma_0 s_x$. With the help of $R_y$ we find the form of the Hamiltonian at one of the two inequivalent $K$ points. All terms, that change sign under $R_y$ will thus be modified by the additional parameter $\tau_z=\pm1$ to mark the difference between
the valleys $K$ and $K'$.
The true time reversal operator equals $\mathcal{T}= - \sigma_0 \tau_x s_y \mathcal{C}$, where $\mathcal{C}$ is the complex conjugation operator \cite{Been} and $\tau_x$ a Pauli matrix corresponding to the valley isospin. Its impact on the Hamiltonian can be written as
\begin{equation*}
\mathcal{T} \mathcal{H}^K(\vec{\mathcal{K}}) \mathcal{T}^{-1}=\mathcal{H}^{K' *}(\xi\vec{\mathcal{K}}),
\end{equation*}
where $\xi=\pm 1$ depending on whether the tensor component $\vec{\mathcal{K}}$ changes sign under time reversal or not. Hence, both operators $R_y$ and $\mathcal{T}$ lead to transformations between the valleys. We now combine both of them to a (new) time-reversal operator within a single valley $\Theta$ (in the spirit of Ref.~\cite{WinklerPa}): $\Theta(R_y)=\mathcal{T} \mathcal{D}(R_y)=\imath s_z \mathcal{C}$. This operator yields an additional symmetry constraint for all terms
\begin{equation}
\Theta \mathcal{H}(\vec{\mathcal{K}}) \Theta^{-1}=s_z \mathcal{H}^{*}(\xi R_y^{-1}\vec{\mathcal{K}})s_z,
\end{equation}
which forces the coefficients $a_6$, $a_8$, $a_9$, and $a_{12}$ in Eq.~(\ref{eq:Ham}) to vanish.
Our result is the low-energy Hamiltonian of silicene near the $K$-points (in first orders in $\vec{k}$ and $E_z$) presented in the basis $(\psi_A \beta, \psi_A \alpha, \psi_B \beta, \psi_B \alpha)^T$. It is given by
\begin{eqnarray}
\mathcal{H}^{K(K')} & = a_1\mathbb{I}+ a_2 \tau_z \sigma_z s_z  +  a
_3 \sigma_z s_0 E_z + a_4 \tau_z \sigma_0 s_z E_z + \\
& a_5 ( \tau_z \sigma_x s_y - \sigma_y s_x) E_z+  a_7 (\tau
_z \sigma_x k_x +\sigma_y k_y) s_0 + \nonumber \\
& a_{10} E_z [\sigma_x(s_x k_y + s_y k_x)+\tau_z\sigma_y(s_x k_x -s_y k_y)] \nonumber+\\
& a_{11} E_z \sigma_0 (s_x k_y - s_y k_x) + a_{13}\sigma_z (s_x k_y - s_y k_x) + \nonumber \\
& a_{16}E_z(\sigma_x k_x +\tau_z\sigma_y k_y)s_z. \nonumber
\end{eqnarray}

\section{Tight-binding model with rectangular unit cell}
\label{sec:AppB}

The terms contributing to the tight-binding Hamiltonian in Eq.~(\ref{eq:HTotal}), in the basis
$\{|\psi_A\rangle, |\psi_B\rangle, |\psi_{A'}\rangle, |\psi_{B'}\rangle\} \times \{\uparrow, \downarrow \}$, take
the explicit form
\begin{equation}\label{eq:AH1}
\footnotesize
H_t=\left(
\begin{array}{cccc}
 0 & t_1 & 0 & t_1' \\
 t_1^* & 0 & t_2' & 0 \\
 0 & t_2'^* & 0 & t_1^* \\
 t_1'^* & 0 & t_1 & 0
\end{array}
\right)\times \mathbb{I}_{(2\times2)},
\end{equation}
\begin{equation}
\footnotesize
 H_{so}=
\left(
\begin{array}{cccccccc}
 s_1 & 0 & 0 & 0 & s_1' & 0 & 0 & 0 \\
 0 & -s_1 & 0 & 0 & 0 & -s_1' & 0 & 0 \\
 0 & 0 & -s_1 & 0 & 0 & 0 & -s_2' & 0 \\
 0 & 0 & 0 & s_1 & 0 & 0 & 0 & s_2' \\
 s_1'^* & 0 & 0 & 0 & s_1 & 0 & 0 & 0 \\
 0 & -s_1'^* & 0 & 0 & 0 & -s_1 & 0 & 0 \\
 0 & 0 & -s_2'^* & 0 & 0 & 0 & -s_1 & 0 \\
 0 & 0 & 0 & s_2'^* & 0 & 0 & 0 & s_1
\end{array}
\right),
\end{equation}
\begin{equation}
\footnotesize
 H_{e}=
\left(
\begin{array}{cccc}
 e_1 & 0 & 0 & 0 \\
 0 & e_1 & 0 & 0 \\
 0 & 0 & -e_1 & 0 \\
 0 & 0 & 0 & -e_1
\end{array}
\right)\times \mathbb{I}_{(2\times2)},
\end{equation}
\begin{equation}
\footnotesize
 H_{e,2}=
\left(
\begin{array}{cccccccc}
 e_2 & 0 & 0 & 0 & e_2' & 0 & 0 & 0 \\
 0 & -e_2 & 0 & 0 & 0 & -e_2' & 0 & 0 \\
 0 & 0 & e_2 & 0 & 0 & 0 & e_3' & 0 \\
 0 & 0 & 0 & -e_2 & 0 & 0 & 0 & -e_3' \\
 e_2'^* & 0 & 0 & 0 & e_2 & 0 & 0 & 0 \\
 0 & -e_2'^* & 0 & 0 & 0 & -e_2 & 0 & 0 \\
 0 & 0 & e_3'^* & 0 & 0 & 0 & e_2 & 0 \\
 0 & 0 & 0 & -e_3'^* & 0 & 0 & 0 & -e_2
\end{array}
\right),
\end{equation}
\begin{equation}
\footnotesize
 H_{\lambda_{r,1}}=
\left(
\begin{array}{cccccccc}
 0 & 0 & 0 & r_1 & 0 & 0 & 0 & r_1' \\
 0 & 0 & r_2 & 0 & 0 & 0 & r_1' & 0 \\
 0 & r_2^* & 0 &  0 & 0 & r_2' & 0 & 0 \\
r_1^* & 0 & 0 &  0 & r_2' & 0 & 0 & 0 \\
 0 & 0 & 0 & r_2'^* & 0 & 0 & 0 & -r_1^* \\
 0 & 0 & r_2'^* & 0 & 0 & 0 & -r_2^* & 0 \\
 0 & r_1'^* & 0 & 0 & 0 & -r_2 & 0 & 0 \\
 r_1'^* & 0 & 0 & 0 &  -r_1 & 0 & 0 & 0
\end{array}
\right),
\end{equation}
\begin{equation}\label{eq:AH6}
\footnotesize
 H_{\lambda_{r,2}}=
\left(
\begin{array}{cccccccc}
 0 & u_1 & 0 & 0 & 0 & u_1' & 0 & 0 \\
 u_1^* & 0 & 0 & 0 & u_2' & 0 & 0 & 0 \\
 0 & 0 & 0 & -u_1 & 0 & 0 & 0 & u_3' \\
 0 & 0 & -u_1^* & 0 & 0 & 0 & u_4' & 0 \\
 0 & u_2'^* & 0 & 0 & 0 & -u_1^* & 0 & 0 \\
 u_1'^* & 0 & 0 & 0 & -u_1 & 0 & 0 & 0 \\
 0 & 0 & 0 & u_4'^* & 0 & 0 & 0 & u_1^* \\
 0 & 0 & u_3'^* & 0 & 0 & 0 & u_1 & 0
\end{array}
\right).
\end{equation}
The matrix elements are given in tables \ref{tab:bigHamCo} and \ref{tab:bigHamCo2}.\\
\begin{table}[htb]
\caption{\label{tab:bigHamCo} Matrix elements in Eqs.~(\ref{eq:AH1}) to (\ref{eq:AH6}) of the diagonal matrix blocks. We use the abbreviations $x\equiv \frac{\sqrt{3} a}{2} k_x$ and $y\equiv \frac{a}{2} k_y$.}
\tiny
\begin{tabular}{@{} llll}
\br
$t_1$ & $2 t \cos(x) (\cos(x) + \imath \sin(x))$ & $u_1$ & $ \frac{8}{3} \imath \lambda_{r,2} \cos(x) \sin(x)$ \\
$s_1$ & $-2 \lambda_{so} \sin(2x) $ & $e_1$ & $\lambda_e E_z $ \\
$e_2$ & $s_1 e_1 \frac{\lambda_{e,2}}{\lambda_{so} \lambda_e}$
& $r_1$ & $\imath \frac{\lambda_{r,1} E_z}{\sqrt{1+\cot^2(\theta)}} (\cos(x) + \imath \sin(x)) (\cos(x) + \sqrt{3} \sin(x))$ \\
$r_2$ & $\imath \frac{\lambda_{r,1} E_z}{\sqrt{1+\cot^2(\theta)}} (\cos(x) + \imath \sin(x)) (\cos(x) - \sqrt{3} \sin(x))$ &  &\\
\br
\end{tabular}
\end{table}
\begin{table}[htb]
\caption{\label{tab:bigHamCo2} Matrix elements in Eqs.~(\ref{eq:AH1}) to (\ref{eq:AH6}) of the non-diagonal matrix blocks. We use the abbreviations $x\equiv \frac{\sqrt{3} a}{2} k_x$ and $y\equiv \frac{a}{2} k_y$.}
\tiny
\begin{tabular}{@{} l p{6.9cm} l  p{6.9cm}}
\br
$t_1'$ & $t (\cos (6 y)+\imath \sin (6 y)) $ & $t_2'$  & $t$ \\
$s_1'$ & $4 \lambda_{so} \sin(x) \cos(3y) \left[\cos \left(x+3y \right)+\imath \sin \left( x+3y \right)\right]$ & $e_2'$ & $s_1' e_1 \frac{\lambda_{e,2}}{\lambda_{so} \lambda_e}$ \\
$e_3'$ & $s_2' e_1 \frac{\lambda_{e,2}}{\lambda_{so} \lambda_e}$ &
$s_2'$ & $4 \lambda_{so} \sin(x) \cos(3y) \left[-\cos \left(x-3y \right)+\imath \sin \left( x-3y \right)\right]$ \\
$r_1'$ & $ \frac{\lambda_{r,1} E_z}{\sqrt{1+\cot^2(\theta)}} (\sin (6y)- \imath \cos (6y))$ & $r_2'$ & $\imath \frac{\lambda_{r,1} E_z}{\sqrt{1+\cot^2(\theta)}}$ \\
$u_1'$ & $\frac{2 \lambda_{r,2}}{3\sqrt{3}}
 \left(\imath \left(\sqrt{3}+3 \imath\right) \sin (x-3 y)+\left(3+\imath \sqrt{3}\right)
   \sin (x+3 y)\right)\times\ \ \ (\cos (x+3 y)+\imath \sin (x+3 y))$ &
$u_2'$ & $\frac{2 \lambda_{r,2}}{3\sqrt{3}} \left(\left(\sqrt{3}-3 \imath\right) \sin (x-3 y)+\left(\sqrt{3}+3 \imath\right) \sin
   (x+3 y)\right)\times\ \ \ (\sin (x+3 y)-\imath \cos (x+3 y))$ \\
$u_3'$ & $\frac{2 \lambda_{r,2}}{3\sqrt{3}} \left(\left(3-\imath \sqrt{3}\right) \sin (x-3 y)+\left(-3-\imath \sqrt{3}\right)
   \sin (x+3 y)\right)\times\ \ \ (\cos (x-3 y)-\imath \sin (x-3 y))$ &
$u_4'$ & $\frac{2 \lambda_{r,2}}{3\sqrt{3}} \left(\left(\sqrt{3}-3 \imath\right) \sin (x-3 y)+\left(\sqrt{3}+3 \imath\right) \sin
   (x+3 y)\right)\times\ \ \ (\sin (x-3 y)+\imath \cos (x-3 y))$ \\
\br
\end{tabular}
\end{table} \\
Note, that coefficients are related to parameters of the group theoretical approach by
\begin{eqnarray*}
& a_1=3\sqrt{3} \lambda_{so}, \hspace{1cm} a_2=\lambda_e, \\
& a_3= 3\sqrt{3} \lambda_{e,2}, \hspace{1cm} a_4=\lambda_{r,1}\ \frac{3}{2} \frac{1}{\sqrt{1+\cot^2(\theta)}}, \\
& a_5= t a \frac{3}{2}, \hspace{1.5cm} a_6=-\lambda_{r,2} a\ \sqrt{3}.
\end{eqnarray*}

\end{document}